\documentclass[format=acmsmall, review=false, screen=true,authorversion]{acmart}

\usepackage{booktabs} 

\usepackage[ruled]{algorithm2e} 

\SetAlFnt{\small}
\SetAlCapFnt{\small}
\SetAlCapNameFnt{\small}
\SetAlCapHSkip{0pt}
\IncMargin{-\parindent}
\usepackage{todonotes}

\setcopyright{acmcopyright}
\acmJournal{TOCHI}
\acmYear{2021} \acmVolume{28} \acmNumber{1} \acmArticle{1} \acmMonth{1} \acmPrice{15.00}\acmDOI{10.1145/3414472}

\begin{document}
\title{Designing Deep Reinforcement Learning for Human Parameter Exploration}

\author{Hugo Scurto}
\affiliation{%
  \institution{STMS Lab, IRCAM--CNRS--Sorbonne Universit{\'e}}
  \streetaddress{1 Place Igor Stravinsky}
  \city{Paris}
  \country{France}}
\email{Hugo.Scurto@ircam.fr}

\author{Bavo Van Kerrebroeck}
\affiliation{%
  \institution{STMS Lab, IRCAM--CNRS--Sorbonne Universit{\'e}}
  \streetaddress{1 Place Igor Stravinsky}
  \city{Paris}
  \country{France}}
\email{Bavo.VanKerrebroeck@ircam.fr}

\author{Baptiste Caramiaux}
\affiliation{%
  \institution{CNRS--LRI, Universit{\'e} Paris-Sud}
  \streetaddress{B\^at 650, Rue Noetzlin, Orsay, F-91400}
  \city{Paris}
  \country{France}}
\email{baptiste.caramiaux@lri.fr}

\author{Fr{\'e}d{\'e}ric Bevilacqua}
\affiliation{%
  \institution{STMS Lab, IRCAM--CNRS--Sorbonne Universit{\'e}}
  \streetaddress{1 Place Igor Stravinsky}
  \city{Paris}
  \country{France}}
\email{Frederic.Bevilacqua@ircam.fr}

\renewcommand{\shortauthors}{H. Scurto et al.}
\renewcommand{\shorttitle}{Designing Deep Reinforcement Learning for Human Parameter Exploration}

\begin{abstract}
Software tools for generating digital sound often present users with high-dimensional, parametric interfaces, that may not facilitate exploration of diverse sound designs. In this paper, we propose to investigate artificial agents using deep reinforcement learning to explore parameter spaces in partnership with users for sound design. We describe a series of user-centred studies to probe the creative benefits of these agents and adapting their design to exploration. Preliminary studies observing users' exploration strategies with parametric interfaces and testing different agent exploration behaviours led to the design of a fully-functioning prototype, called Co-Explorer, that we evaluated in a workshop with professional sound designers. We found that the Co-Explorer enables a novel creative workflow centred on human-machine partnership, which has been positively received by practitioners. We also highlight varied user exploration behaviors throughout partnering with our system. Finally, we frame design guidelines for enabling such co-exploration workflow in creative digital applications.
\end{abstract}

%
%



\maketitle

\section{Introduction}\label{sec:intro}

Reinforcement learning defines a computational framework for the interaction between a learning agent and its environment \cite{niv2009reinforcement}. The framework provides a basis for algorithms that learn an optimal behaviour in relation to the goal of a task \cite{Sutton2011}. For example, reinforcement learning was recently used to learn to play the game of Go, simulating thousands of agent self-play games based on human expert games \cite{silver2016mastering}. The algorithm, called \textit{deep reinforcement learning}, leveraged advances in deep neural networks to tackle learning of a behaviour in high-dimensional spaces \cite{mnih2015human}. The autonomous abilities of deep reinforcement learning agents let machine learning researchers foresee prominent applications in several domains, such as transportation, healthcare, or finance \cite{li2018deep}.

Yet, one important current challenge for real-world applications is the ability for reinforcement learning agents to learn from interaction with human users. The so-called \textit{interactive reinforcement learning} framework has been shown to hold great potential to build autonomous systems that are centered on human users \cite{amershi2014power}, such as teachable and social robots \cite{thomaz2008teachable}, or assistive search engines \cite{athukorala2016beyond}. From a machine learning perspective, the main challenge lies in learning an optimal behaviour from small, non-stationary amounts of human data \cite{knox2009interactively}. From a human-computer interaction perspective, an important challenge consists in supporting human appropriation of agents' autonomous behaviours in relation to complex human tasks \cite{stumpf2009interacting}.

Our interest lies in investigating interactive reinforcement learning for human creative tasks, where a goal might not be well-defined by human users a priori \cite{resnick2005design}. One such case of a human creative task is \textit{exploration} \cite{hart2017creative}. Exploration consists in trying different solutions to address a problem, encouraging the co-evolution of the solution and the problem itself \cite{dorst2001creativity}. For example, designers may produce several sketches of a product to ideate the features of its final design, or test several parameter combinations of a software tool to create alternative designs in the case where the product has a digital form. The creative, human-centred, use case of exploration thus fundamentally differs from standard, machine-centred, reinforcement learning use cases, where a problem is implicitly defined as a goal behaviour, before the agent actually learns to find an optimal solution \cite{Sutton2011}. Here, one could expect that humans would prefer agent autonomous behaviours---provoking surprise and discovery along exploration---over the learning of one optimal solution---forcing human users to teach agents one optimal behaviour.

In this paper, we aim at designing an interactive reinforcement learning system supporting human creative exploration. This question is addressed in the application domain of sound design, where practitioners typically face the challenge of exploring high-dimensional, parametric sound spaces. We propose a user-centred design approach with expert sound designers to steer the design of such a system and conceptualize exploration within this context. We conducted two case studies to evaluate two prototypes that we developed. The final prototype specifically designed deep reinforcement learning to foster human exploration. Specifically, it employs reinforcement learning as an algorithmic approach to dynamically suggest different sounds to users based on their feedback data---thus possibly contributing to their exploration process. Therefore, contrary to typical reinforcement learning-based tools, it does not aim at creating fully-trained agents that could be optimally reused by users across different sessions. Our overall proposed methodology thus radically differs from standard reinforcement learning approaches.

Our findings led to contributions at several levels. On the conceptual side, we were able to characterize different user approaches to exploration, and to what we have called \textit{co-exploration}---exploration in cooperation with an interactive reinforcement learning agent. These range from analytical to spontaneous in the former case, and from user- to agent-as-leader in the latter. On the technical side, a user-centered approach let us adapt a deep reinforcement learning algorithm to the case of co-exploration in high-dimensional parameter spaces. This notably required creating additional interaction modalities to user reinforcement, jointly with an autonomous exploration behaviour for the reinforcement learning agent. Crucially, our qualitative results suggest that the resulting agent---especially its non-optimal nor predictable behaviour---may be well experienced by human users leading parameter exploration, in more creative ways than random conditions. Lastly, on the design side, we extracted a set of important challenges that we deem critical for joint HCI and machine learning design in creative applications. These include: (1) engaging users with machine learning, (2) foster diverse creative processes, and (3) steer users outside comfort zones.

\section{Related Work}

In this section, we review related work on machine learning in the field of Human-Computer Interaction, encompassing creativity support tools, interactive machine learning, and interactive reinforcement learning, with a focus on human exploration.

\subsection{Creativity Support Tools}

Human creative exploration of multidimensional parameter spaces has long been studied in the NIME community (acronym for New Interfaces for Musical Expression, originally emerging as a workshop at CHI \cite{poupyrev2001new}). Techniques for mapping input user parameters (e.g., gestural control data) to output multidimensional parameters (e.g., sound synthesis parameters) were developed to support exploration through new digital music instruments \cite{hunt2000mapping,hunt2002mapping}. Perceptual descriptors of sound enabled to reduce dimensionality of sound spaces, thus providing users with more intuitive interfaces to explore sound spaces \cite{schwarz2009sound}. Yet, besides a few exceptions \cite{fiebrink2010toward}, the creative process of human users leading parameter exploration with computer music software has still been hardy investigated from a research perspective.

Creativity support tools have long focused on human exploration as a central process to human creative work \cite{shneiderman2007creativity}. Design guidelines for supporting exploration were developed, which include aiming at simple interfaces for appropriating the tool and getting into sophisticated interaction more easily \cite{Dix2007}. Flexible interaction modalities that can adapt to users' very own styles of thinking and creating may also be required \cite{resnick2005design}. In particular, parameter space exploration remains a current challenge for HCI research \cite{cartwright2014mixploration}. Recently, creativity-oriented HCI researchers underlined the need to move toward interdisciplinary research collaborations \cite{pedersen2018twenty}.

Machine learning was in this sense examined for its implications in design \cite{koch2017design} and identified as an opportunity for user experience \cite{dove2017ux,yang2018mapping,yang2018investigating}. Yet, a large body of work in the machine learning research community has so far focused on constructing autonomous algorithms learning creative behaviour from large amounts of impersonal data---falling under the name of computational creativity \cite{wiggins2006preliminary}. While this have allowed the building of powerful tools and models for creation, one may be concerned in the question of how to include human users in the design of such models to support human-computer co-creation \cite{kantosalo2014isolation}.

Davis et al. proposed a model of creativity that explicitly considers the computer as an enactive entity \cite{davis2014building}. They notably stressed the potential of combining creativity support tools with computational creativity to enrich a collaborative process between the user and the computer \cite{davis2014building}. The Drawing Apprentice, a co-creative agent that improvizes in real-time with users as they draw, illustrates their approach \cite{davis2016empirically}. While their user study confirms the conceptual potential of building such artistic computer colleagues, its technical implementation remains specific to the use case at stake---\textit{e.g.}, drawing. We propose to jointly design a conceptual and technical framework that could be easily transferable to other application domains---potentially realizing general mixed-initiative co-creativity \cite{horvitz1999principles,yannakakis2014mixed}.

\subsection{Interactive Machine Learning}

Interactive machine learning \cite{fails2003interactive} allows human users to build customized models by providing their own data examples---typically a few of them. Not only users can customize training examples, but they are also allowed to directly manipulate algorithm parameters \cite{kapoor2010interactive,wong2011end}, as well as to receive information on the model's internal state \cite{amershi2015modeltracker, patel2011using}. Applications in HCI cover a wide range of tasks, such as handwriting analysis \cite{shilman2006cuetip}, recommender systems \cite{amershi2012regroup}, or prioritising notifications \cite{amershi2011cuet}. Interactive machine learning mainly builds on supervised learning, which defines a computational framework for the learning of complex input-output models based on example input-output pairs. The ``human-in-the-loop'' approach to supervised learning critically differs from the computational creativity approach, which typically relies on huge, impersonal databases to learn models \cite{gillies2016human}.

Interactive machine learning is one such example of a generic framework for human-computer co-creation \cite{amershi2014power}. The technical framework was successfully applied across several creative domains, such as movement interaction design \cite{zamborlin2014fluid,francoise2018motion,gillies2019understanding}, web page design \cite{kumar2011bricolage} or video games \cite{kleinsmith2013customizing}. Specifically, research studying users building customized gestural controllers for music brought insight on the creative benefits of interacting with machine learning \cite{Fiebrink2011}. Not only were users able to accomplish their design goal---\textit{e.g.}, demonstrating a given gesture input for controlling a given sound parameter output---, but they also managed to explore and rapidly prototype alternative designs by structuring and changing training examples \cite{fiebrink2010toward}. These patterns were reproduced by novice users who gained accessibility using examples rather than raw parameters as input \cite{katan2015using}. The algorithms' sometimes surprising and unexpected outcomes favoured creative thinking and sense of partnership in human users \cite{fiebrink2016machine}.

Typical workflows in interactive machine learning tend to iterate on designing training examples that are built from a priori representative features of the input space to support exploration. Yet, in some creative tasks where a problem definition may be found only by arriving at a solution \cite{dorst2001creativity,rittel1972planning}, it might be unfeasible for users to define, a priori, such representative features of the final design \cite{katan2015using}. Other approaches proposed methods to release such contraints, for example by exploring alternative machine learning designs by only defining the limits of some parameter space \cite{scurto2016grab}. We propose to further investigate machine learning frameworks able to iteratively learn from other user input modalities, and explicitly considering mixed-initiative workflows, where systems autonomously adapt to users \cite{deterding2017mixed}. As reviewed in the next section, using interactive reinforcement learning offers such perspectives.

\subsection{Interactive Reinforcement Learning}

Interactive reinforcement learning defines a computational framework for the interaction between a learning agent, a human user, and an environment \cite{amershi2014power}. Specifically, users can communicate positive or negative \textit{feedback} to the agent, in the form of a numerical reward signal, to teach it which action to take when in a certain environment state. The agent is thus able to adapt its behaviour to users, while remaining capable of behaving autonomously in its environment.

While user feedback has been used as input modality for applications in information retrieval \cite{zhou2003relevance}, recommender systems \cite{koren2009matrix}, or affective computing \cite{liu2009dynamic}, it was often included in algorithmic frameworks relying on pre-established, rule-based methods to provide users with adaptive behaviour. The data-driven abilities of reinforcement learning, in contrast, offers promising perspectives for open-ended, interactive applications that are centered on human users. In this sense, interactive reinforcement learning relies on small, user-specific data sets, which contrasts with the large, crowdsourced data sets used in creative applications in semantic editing \cite{yumer2015semantic,koyama2016computational,desai2019geppetto}. Lastly, interactive approaches to reinforcement learning focuses on exploring agent actions based on human feedback on actions, which contrasts with the focus on optimising one parametric state based on user feedback over states---as used in Bayesian Optimisation \cite{brochu2010bayesian,liu2017bignav} or multi-armed bandits \cite{lomas2016interface}. 

Interactive reinforcement learning has been recently applied in HCI \cite{Ruotsalo2014}, with promising applications in exploratory search \cite{Glowacka2013,Athukorala2016} and adaptive environments \cite{frenoy2016adaptive,rajaonarivo2017inline}. Integrating user feedback in reinforcement learning algorithms is computationally feasible \cite{stumpf2007toward}, helps agents learn better \cite{knox2009interactively}, can make data-driven design more accessible \cite{lomas2016interface}, and holds potential for rich human-computer collaboration \cite{stumpf2009interacting}. Applications in Human-Robot Interaction informed on how humans may give feedback to learning agents \cite{thomaz2008teachable}, and showed potential for enabling human-robot co-creativity \cite{fitzgerald2017human}. Recently, reinforcement learning has witnessed a rise in popularity thanks to advances in deep neural networks \cite{mnih2015human}. Powerful models including user feedback have been developed for high-dimensional parameter spaces \cite{christiano2017deep,warnell2017deep}. Design researchers have identified reinforcement learning as a promising prospective technique to improve human-machine ``joint cognitive and creative capacity'' \cite{koch2018group}.

We believe that interactive reinforcement learning---especially deep reinforcement learning---holds great potential for supporting creative tasks---especially exploration of high-dimensional parameter spaces. First, its computational framework, constituted by environment states, agent actions, and user feedback, remains fully generic \cite{Sutton2011}, and thus potentially allows the design of generic interaction modalities transferrable to different application domains. Second, the action-oriented, autonomous exploration behaviour intrinsic to reinforcement learning algorithms may be exploited to build a novel creative mixed-initiative paradigm, where the user and the agent would cooperate by taking actions that are ``neither fully aligned nor fully in conflict'' \cite{crandall2018cooperating}. Finally, we consider that user feedback could be a relevant input modality in the case of exploration, notably for expressing on-the-fly, arbitrary preferences toward imminent modifications, as opposed to representative examples. As previously stated, this requires investigating a somewhat unconventional use of reinforcement learning: if previous works employed user feedback to teach agents an optimal behavior in relation to a task's goal, it is less obvious whether such an optimal behavior may be well-defined---or even exists---for human users performing exploration.

\section{General Approach}

In this section, we describe the general approach of our paper, applying interactive reinforcement learning for human parameter space exploration in the creative domain of sound design.

\subsection{Problem Formulation}

In this paper, we seek to address the following general research question: \textit{How to design reinforcement learning to support human parameter exploration?} While many formulations of reinforcement learning (RL) could be imagined to address this question, our approach focused on an interactive use case of reinforcement learning, where feedback data is provided in real-time by a human user. We hypothesize that such formulation may be of interest for users leading parameter exploration.

We investigate the reinforcement learning problem in the context of classical sequential decision making. Let $\mathcal{S}=\{S\}$ denote the state space constituted by all possible parameter configurations $S=(s_1,...,s_n)$ reachable by the agent, with $n$ being the number of parameters, and $s_i \in \lbrack s_{min},s_{max}\rbrack$ being the value of the $i^{th}$ parameter living in some bounded numerical range. Let $\mathcal{A}(S)=\{A\}$ denote the corresponding action space as moving up or down one of the $n$ parameters by one step $a_i$, except when the selected parameter equals one boundary value. As the agent selects actions and iteratively acts on parameters one by one, we assume that a human observes the state-action path and interactively provides positive of negative feedback, $R$, to the agent. The agent's goal is to maximize user feedback, which it does by learning a mapping between state and actions defined from user feedback. In order to allow such real-time human interaction with the currently-learning agent, we only consider on-policy settings of reinforcement learning---as opposed to off-policy settings, which separates a behaviour policy for real-time environment exploration from an estimation policy that learns from reinforcement \cite{Sutton2011}. Designing reinforcement learning for human parameter exploration thus consists in understanding what interactive mechanisms may be relevant for the agent to support human exploration.

Learning action values instead of state values constitutes the main reason for treating the task as an RL problem. By definition, human parameter exploration has a strong dependence between current state and action to the next state. It is precisely there that lies our rationale behind RL: to account for human actions in parameter state exploration. Conversely, Bayesian Optimisation (BO) algorithms do not take into account actions leading to states when exploring parameter states. As such, BO may be relevant for parameter search---e.g., optimization of a parametric state based on user feedback over states---, but not for parameter exploration---e.g., trial of parameter actions based on user feedback over actions. Contextual Bandits, on the other hand, only afford learning of action values from a small, discrete set of states. In our case study, we will show that learning in continuous state-action spaces is required to tackle parameter space exploration, and that reinforcement learning---especially its deep learning extension---suits this task. We will also observe that the interactive reinforcement learning formulation enables users to explore parameter spaces by providing relatively small amounts of feedback data---an average of 235 feedback data during a typical session, with a large standard deviation corresponding to diverse user feedback behaviours. This is in contrast with the large datasets often required in more standard RL applications to continuous action spaces.

\subsection{Application Domain}

Sound design is an exemplary application domain for studying exploration---taking iterative actions and multiple steps to move from an ill-formed idea to a concrete realization \cite{garcia2012interactive}. Sonic exploration tasks can take myriad of forms: for example, composers explore various sketches of their musical ideas to write a final score; musicians explore different playing modes to shape an instrument's tone; sound designers explore several digital audio parameters to create unheard-of sounds \cite{monache2010toolkit,delle2018embodied}. 

\begin{figure}[h]
  \includegraphics[width=.75\columnwidth]{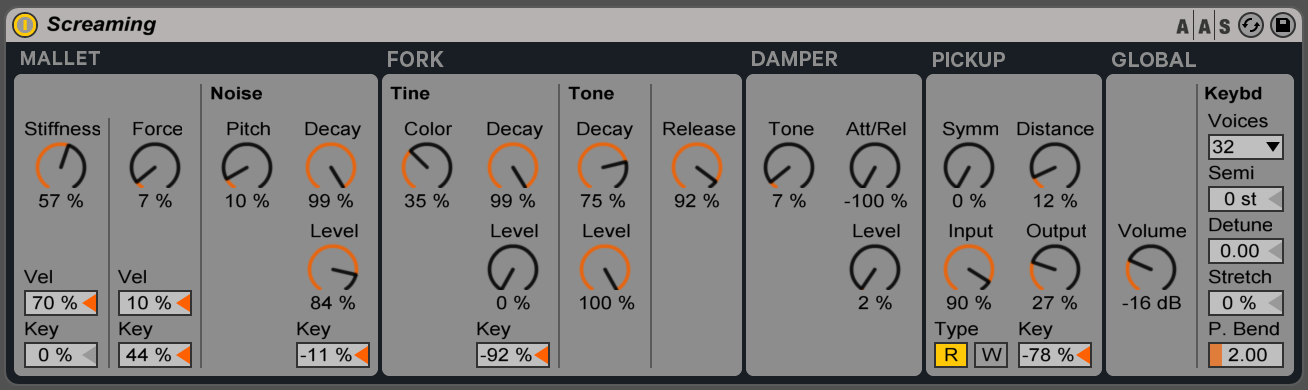}
  \caption[Caption]{A typical VST interface in \textit{Ableton Live} (sound design software), containing many technical parameters.}
  \label{fig:standard}
\end{figure}

Most of today's digital commercial tools for sound synthesis, named Virtual Studio Technology (VST, see Fig. \ref{fig:standard}), still rely on complex interfaces using tens of technical parameters as inputs. These parameters often relate to the underlying algorithms that support sound synthesis, preventing users from establishing a direct perceptual relationship with the sound output. To that one may add the exponential number of parameter combinations, called presets, that eventually correspond to given sound designs. It is arguable that these interfaces may not be the best to support human exploration: as the perceptual outcome of acting on a given parameter may rapidly become unpredictable, they may hinder user appropriation \cite{resnick2005design,shneiderman2007creativity}.

By formalizing human parameter space exploration as an interactive reinforcement learning problem, we seek to tackle both issues at once. First, human navigation in high-dimensional parameter spaces may be facilitated by the reinforcement learning computational framework, made of sequences of states, actions, and rewards. Second, human creativity may be stimulated by the action-oriented, autonomous behaviour of reinforcement learning algorithms, suggesting other directions or design solutions to users along exploration.

\subsection{Method}

We adopted a user-centered approach to lead joint conceptual and technical work on interactive reinforcement learning for parameter space exploration. We decided to work with expert sound designers to get feedback on the creative task of parameter space exploration as it is led by specialized users. This qualitative understanding would in turn inform the design of interactive reinforcement learning in its application to sound design. Two design iterations---a pilot study and an evaluation workshop---were conducted over the course of our research. Two prototypes were designed and developed---one initial reinforcement learning prototype, and the \textit{Co-Explorer}, our final deep reinforcement learning prototype. The process thus includes sequentially: 

\begin{itemize}
\item Prototype 1: Implementing a reinforcement learning algorithm that learns to explore sound parameter spaces from binary human feedback
\item Pilot study, Part 1: Observing and interviewing participants exploring sound spaces with standard parametric interfaces
\item Pilot study, Part 2: Observing and interviewing participants using our initial reinforcement learning prototype to explore a sound space
\item Prototype 2: Designing deep reinforcement learning in response to design ideas suggested by our pilot study, implementing it in the \textit{Co-Explorer}
\item Workshop, Part 1: Observing and discussing with participants using the Co-Explorer, our final prototype, in an exploration task related to discovery
\item Workshop, Part 2: Observing and discussing with participants appropriating the Co-Explorer, our final prototype, in an exploration task related to creation
\end{itemize}

We worked with a total of 14 users (5 women, 9 men; all French) through the series of activities. From the 14 total, there were 2 who took part in all of the activities listed below, to testify of our prototype's improvements. Our users covered different areas of expertise in sound design and ranged from sound designers, composers, musicians, and artists to music researchers and teachers. Thus, they were not all constrained to one working methodology, one sonic practice or one application domain. Our motivation was to sample diverse approaches to exploration that sound design may provoke, in order to design a flexible reinforcement learning algorithm that may suit a variety of users' working styles \cite{resnick2005design}.

\section{Pilot Study}

We organized a one-day pilot study with four of our expert participants. The aims of this pilot study were to: Observe approaches to exploration in standard parametric interfaces; Identify problems users experience; Introduce the reinforcement learning technology in the form of a prototype; Brainstorm ideas and possible breakdowns.

The study was divided in two parts: (1) parametric interface exploration, then (2) interactive reinforcement learning-based exploration. We conducted individual semi-structured interviews at the end of each part, having each participant do the study one by one. This structure was intended to bring each participant to become aware of their subjective experience of exploration \cite{petitmengin2006describing}. Our intention was to open up discussions and let participants suggest design ideas about interactive reinforcement learning, rather than testing different algorithmic conditions in a controlled, experimental setup. We spent an average of 2 hours with each of our four participants, who covered different expertise in sound design (composition, sound design, interaction design, research).

\subsection{Part 1: Parametric Interfaces}

\subsubsection{Procedure}

\begin{figure}[h]
  \includegraphics[width=.875\columnwidth]{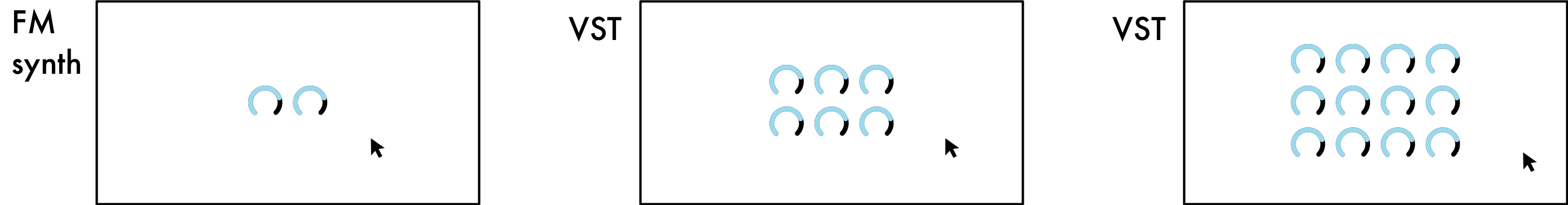}
  \caption{Schematic view of the three parametric interfaces.}
  \label{fig:param}\vspace{-0.25cm}
\end{figure}

In the first part of the study, participants were asked to find and create a sound preset of their choice using three different parametric interfaces with different number of parameters (respectively 2, 6, and 12, see Fig. \ref{fig:param}). No reinforcement learning agent was used. We linked each interface to a different sound synthesis space (respectively created using FM synthesis\footnote{\textit{Frequency Modulation synthesis} (a classic algorithmic method for sound synthesis \cite{chowning1973synthesis}).}, and one commercial VST from which we selected 6, then 12, parameters). Our goal was to investigate how the number of parameters on an interface might influence exploration of large perceptual spaces. As such, we used the FM synth because of the large perceptual space it offers relying on only two parameters.

Sound was synthesized continuously; participants' actions were limited to move the knobs using the mouse to explore the design space offered by all possible combinations. While we agree that tangible interfaces are extensively used by professional sound designers, we underline that the mouse remains used in other creative domains where real-time multi-dimensional control is needed (e.g., graphic design). As such, we decided to use the mouse for this pilot experiment to study the general task of parameter space exploration. Knobs' technical names were hidden to test the generic effect of parameter dimensionality in interface exploration, and avoid any biases due to user knowledge of parameter function (which typically occur with labelled knobs). Interface order was randomized; we let participants spend as much time as they wanted on each interface to let them explore the spaces freely.

\subsubsection{Analysis}

We were interested in observing potential user strategies in parameter space exploration. We thus logged parameter temporal evolution during the task. It consists in an $n$-dimensional vector, with $n$ being the number of parameters (respectively 2, 6, then 12). Sample rate was set to 100 ms, which is a standard value for interaction with sound and musical interfaces \cite{jorda2005digital}. We used Max/MSP\footnote{https://cycling74.com/products/max/} and the MuBu\footnote{https://forum.ircam.fr/projects/detail/mubu/} library to track user actions on parameters and record their evolutions. We used structured observation to study participants' interviews. This method was meant to provide a thorough qualitative analysis on user exploration strategies.

\subsubsection{Observations}

\begin{figure}[h]
  \includegraphics[width=.625\columnwidth]{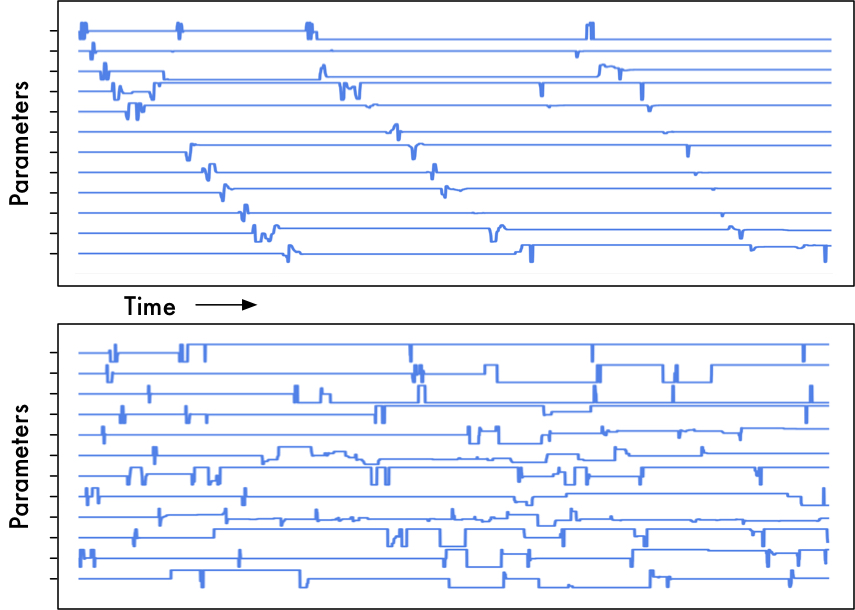}
  \caption{Two user exploration strategies with a 12-dimensional parametric interface: Analytical (top) vs. spontaneous (bottom).}
  \label{fig:exploration}\vspace{-0.25cm}
\end{figure}

Qualitative analysis of parameter temporal evolution let us observe a continuum of approaches to parametric interface exploration. We call the first extremity of this continuum \textbf{analytical exploration}: this involves actioning each of the knobs one after the other over their full range. The second is called \textbf{spontaneous exploration}: this involves making random actions on the knobs. Figure \ref{fig:exploration} shows examples for each of these two approaches. One participant was consistently analytical over the three interfaces; one was consistently spontaneous over the three. The two others combined both approaches over the three interfaces.

Interview analysis let us map these approaches to different subgoals in exploration. The analytical approach concerns exploration of the interface at a parameter level: \textit{``The strategy is to test [each knob] one by one to try to grasp what they do''}, one participant said. The goal of exploration is then related to building a mental map of the parameters to learn how to navigate in the design space. The spontaneous approach concerns exploration of the design space at a creative level: \textit{``I moved the knobs more brutally and as a result of serendipity I came across into something different, that I preferred for other reasons...''}, another participant said. The goal of exploration is then related to discovering new parameter states leading to inspiring parts of the design space.

Discovery is critical to parameter space exploration. \textit{``Once [the knobs] are isolated, you let yourself wander a bit more...''}, one participant analysed. Surprise is also important: \textit{``To explore is to be in a mental state in which you do not aim at something precise''}, one participant said. Interestingly, we observed that participants often used words related to perceptual aspects rather than technical parameters. \textit{``I like when you can get a sound that is... um... Consistent, like, coherent. And at the same time, being able to twist in many different ways. This stimulates imagination, often''}, one participant said. Two participants mentioned that forgetting the parametric interface may be enjoyable in this sense: \textit{``I appreciate an interface that does not indicate [...], that has you go back into sound, so that you are not here reading things, looking at symbols...''}, one participant said. 

All participants reported being hindered in their exploration by the parameter inputs of the three interfaces. As expected, the more parameters the interface contained, the larger the design space was, and the harder it was to learn the interface. \textit{``For me, the most important difficulty is to manage to effectively organise all things to be able to re-use them.''}, one participant said. Time must be spent to first understand, then to memorize the role of parameters, taking into account that their role might change along the path of exploration. This hampers participants' motivation, often restraining themselves to a subspace of the whole design space offered by the tool: \textit{``after a while I was fed up, so I threw out some parameters''}, one participant said about the 12-knob interface.

Participants discussed the limitations encountered in the study in light of their real-world practice with commercial interfaces. Two participants mentioned using automation functions to support parameter space exploration. Such functions include randomizing parameter values, automating parameter modification over time, or creating new control parameters that \textit{``speak more to your sensibility, to your ears, than to what happens in the algorithm''}, to cite one of the participants. Two participants also use factory presets to start exploration: \textit{``I think that in some interfaces they are pretty well conceived for giving you the basis of a design space. Then it's up to you to find what parameters to move''}, one participant said. Two participants said that the graphical user interfaces, including parameter names, knob disposition, and visual feedback on sound, may help them manage to lead exploration of large parameter spaces.

\subsection{Part 2: RL Agent Prototype}

Results in first part let us identify different user approaches to parametric interface exploration, as well as different problems encountered in high-dimensional parameter spaces. In the second part, we were interested in having participants test the reinforcement learning technology in order to scope design ideas and possible breakthroughs in relation to exploration.

\subsubsection{Implementation}\label{subsubsec:proto}

\begin{figure}[h]
  \includegraphics[width=.625\columnwidth]{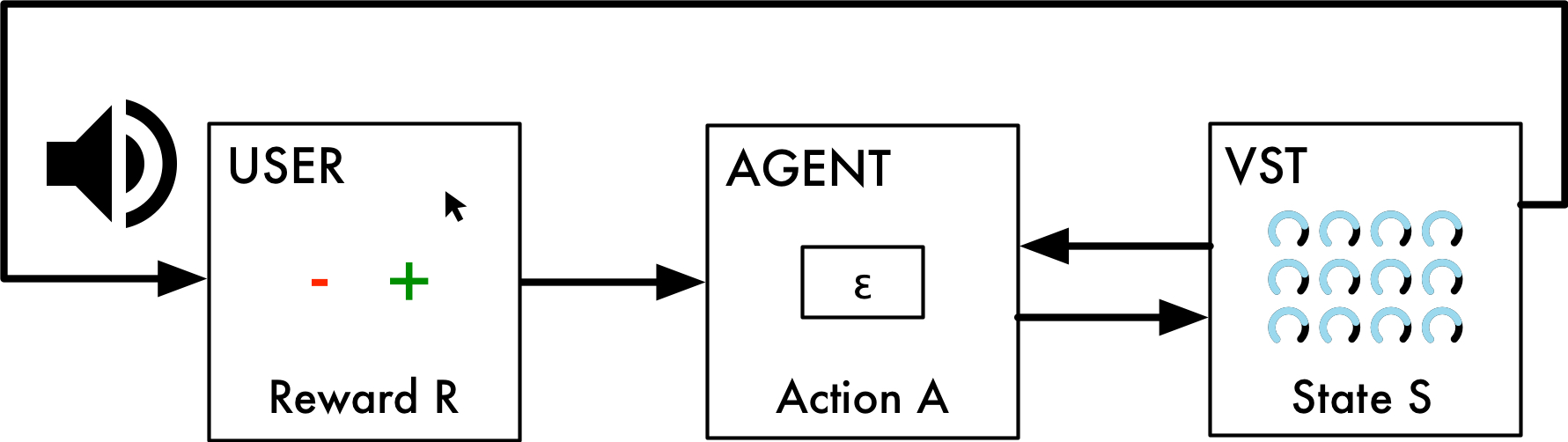}
  \caption{Our RL agent prototype. Users can only provide feedback to the agent, which acts on hidden VST parameters.}
  \label{fig:RL}
\end{figure}

We implemented an initial prototype for our pilot study, that we propose to call ``RL agent'' for concision purposes. The prototype lets users navigate through different sounds by only communicating positive or negative feedback to a reinforcement learning agent. The agent learns from feedback how to act on the underlying synthesis parameters in lieu of users (see Fig. \ref{fig:RL}). Formally, the environment is constituted by the VST parameters, and the agent iteratively acts on them. Computationally, we considered the state space $\mathcal{S}=\{S\}$ constituted by all possible parameter configurations $S=(s_1,...,s_n)$, with $n$ being the number of parameters, and $s_i \in \lbrack s_{min},s_{max}\rbrack$ being the value of the $i^{th}$ parameter living in some bounded numerical range (for example, $s_i$ can control the level of noise normalized between 0 and 1). We defined the corresponding action space $\mathcal{A}(S)=\{A\}$ as moving up or down one of the $n$ parameters by one step $a_i$, except when the selected parameter equals one boundary value:

\begin{equation*}
A(S)=
\begin{cases}
    \pm a_i & \text{for } s_i \in \rbrack s_{min},s_{max}\lbrack\\
    +a_i& \text{for } s_i = s_{min}\\
    -a_i & \text{for } s_i = s_{max}\\
\end{cases}
\end{equation*}

An $\varepsilon$-greedy method defines the autonomous exploration behaviour policy of the agent---how it may take actions by exploiting its accumulated feedback while still exploring unvisited state-action pairs \cite{Sutton2011}. From a given state, it consists in having the agent take the best action with probability $\varepsilon$, and reciprocally, take a random action with probability $1-\varepsilon$. For example, $\varepsilon=1$ would configure an always exploiting agent---\textit{i.e.,} always taking the best actions based on accumulated feedback---, while $\varepsilon=0$ would configure an always exploring agent---\textit{i.e.,} never taking into account the received feedback. Our purpose in this study was to examine whether different exploration-exploitation trade-offs could map to different user approaches to exploration. Finally, we propose that the user would be responsible for generating feedback. We directly mapped user feedback to the environmental reward signal $R$ associated with a given state-action pair $(S,A)$. The resulting formalization---where an agent takes actions that modify the environment's state based on feedback received from a user---defines a generic interactive reinforcement learning problem.

We implemented \textit{Sarsa}, which is a standard algorithm to learn how to act in many different environment state, \textit{i.e.,} for each given parameter configuration \cite{Sutton2011}. It differs from multi-armed bandits, which learns how to act in one unique environment state \cite{lomas2016interface}. Importantly, as evoked in Section \ref{sec:intro}, \textit{Sarsa} was designed to learn one optimal behaviour in relation to the goal of a task. Our purpose in this study was to scope the pros and cons of such a standard reinforcement learning algorithm for human exploration tasks, judging how it may influence user experience, and framing how it may be adapted by design to support exploration. The convergence of the \textit{Sarsa} algorithm in an interactive setup where users provide feedback was evaluated in a complementary work \cite{scurto2018perceiving}.

We used the largest VST-based 12-parameter space of the first part ($n=12$) as the environment of our prototype. Because \textit{Sarsa} is defined on discrete state spaces, each parameter range was discretized in three normalized levels ($s_i \in \{0,0.5,1\},a_i = 0.5; 0 \leq i \leq n$). Our goal was to investigate how the RL agent could help exploration of large perceptual spaces. As such, we opted for more (twelve) parameters and less (three) discrete levels to design the largest perceptual space suitable for our RL-agent prototype. Although this would have been a design flaw in a perceptual experiment on typical VSTs, this allowed for obvious perceptual changes, which was required to investigate feedback-based interaction with a large variety of sounds.

\subsubsection{Procedure}

\begin{figure}[h]
  \includegraphics[width=0.625\columnwidth]{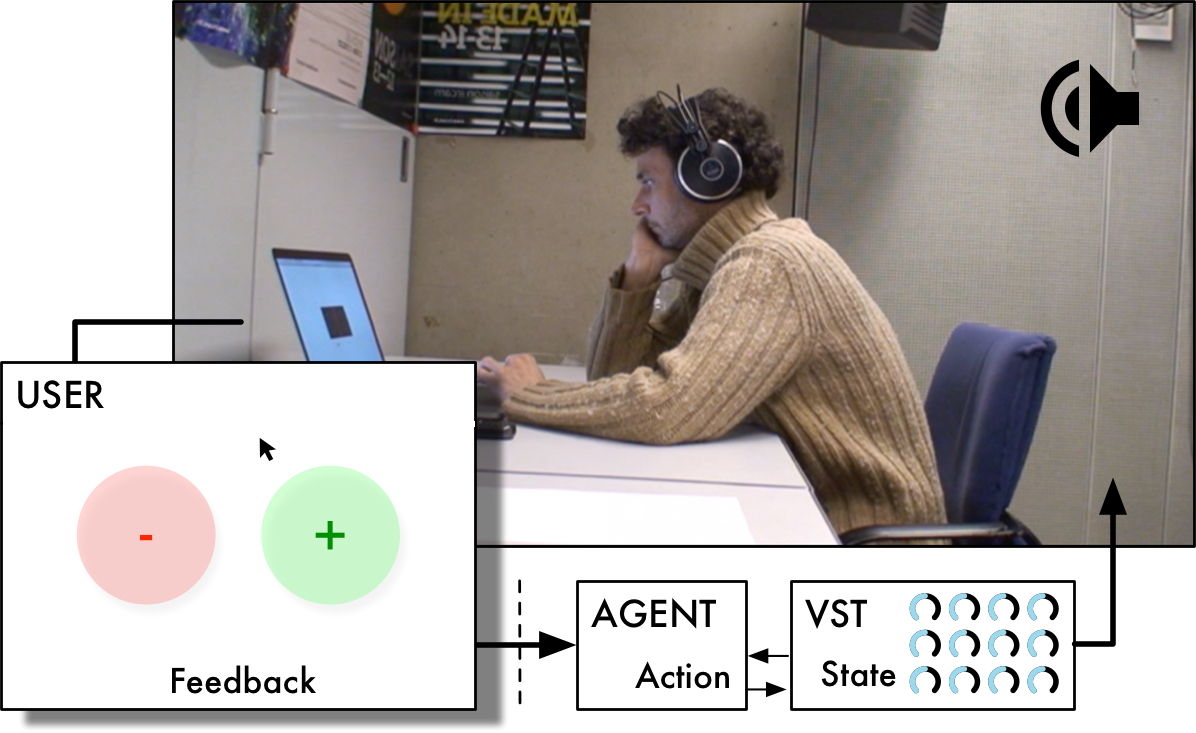}
  \caption{One of our four participants using a two-button interface to communicate binary feedback to the RL agent prototype in the pilot study.}
  \label{fig:study}
\end{figure}

Our participants were asked to find and create a sound preset of their choice by communicating feedback to three different agents with different exploration behaviours (respectively $\varepsilon=0$; $\varepsilon=1$; and $\varepsilon=0.5$). Sound was synthesized continuously, in a sequential workflow driven by the agents' algorithmic functioning. At step $t$, participants could listen to a synthesized sound, and give positive or negative feedback by clicking on a two-button interface (Fig. \ref{fig:study}). This would have the agent take an action on hidden VST parameters, modify the environment's state, and synthesize a new sound at step $t+1$. Participants were only told to give positive feedback when the agent took an action getting them closer to a sound that they enjoy, and negative feedback when it moves away from it. They were not explained the agent's internal functioning, nor the differences between the three agents. Each session started with a fully-untrained agent. The starting state for $t=0$ was randomly selected. Agent order was randomized; we asked participants to spend between 5 and 10 minutes with each.

\subsubsection{Analysis} We logged all participant actions in the graphical user interface. It consisted in timed onsets for positive feedback on the one hand, and negative feedback on the other hand. We also logged parameter temporal evolution to observe how the RL agent would act on parameters following user feedback. We used structured observation to study participants' interviews and discussions led at the end of the pilot study.

\subsubsection{Reactions}\label{subsubsec:reactions}

All participants reported forgetting synthesis parameters to focus on the generated sound. The simplicity and straightforwardness of the new interface benefited their exploration.  \textit{``There's always this sensation that finally you are more focused on listening to the sound itself rather than trying to understand the technology that you have under your hands, which is really great, yeah, this is really great''}, one participant said. Participants went on voluntarily for 5.9 minutes with each of the agents on average ($\sigma = 1.3$). 

The computational framework defined by reinforcement learning was well understood by all participants. \textit{``There's somewhat a good exploration design [sic], because it does a bit what you do [with the parametric interface], you move a thing, you move another thing...''}, one participant said. All participants enjoyed following agents' exploration behaviours, mentioning a playful aspect that may be useful for serendipity. Three participants in turn adapted their exploration to that of the agent: \textit{``you convince yourself that the machine helps you, maybe you convince yourself that it is better... and after you go on exploring in relation to this''}, one participant said. Interestingly, one participant that was skeptical about partnering with a computer changed his mind interacting with the RL agent: \textit{``We are all different, so are they''}, he commented, not without a touch of humor.

\subsubsection{Uses of Feedback}

Descriptive statistics informed on how participants used the feedback channel. Three participants gave feedback every 2.6 seconds on average ($\sigma = 0.4$), globally balancing positive with negative (average of 44.8\% positive, $\sigma = 0.02$). The fourth participant gave feedback every 0.9 seconds on average ($\sigma = 0.07$) which was mostly negative (average of 17.2\% positive, $\sigma = 0.02$). All participants reappropriated the feedback channel, quickly transgressing the task's instructions toward the two-button interface to fulfill their purposes. One participant used feedback to explore agents' possible behaviors: \textit{``Sometimes you click on the other button, like, to see if it will change something, [...] without any justification at all''}, he commented. Another used the `-' button to tell the agent to \textit{``change sound''}. Two participants also noticed the difference between feedback on sound itself, and feedback on the agent's behavior: \textit{``there's the `I don't like' compared to the sound generated before, and the `I don't like it at all', you see''}, one of them said.

\subsubsection{Breakdowns}

Rapidly, though, participants got frustrated interacting with the RL agent. All participants judged that agents did not always reacted properly to their feedback, and were leading exploration at the expense of them: \textit{``sometimes you tell `I don't like', `I don't like', `I don't like', but it keeps straight into it! (laughs)''}, one participant said. Contrary to what we expected, participants did not expressed a strong preference for any of the three tested agents. Only one participant noticed the randomness of the exploring agent, while the three other participants could not distinguish the three agents. This may be caused by the fact that the \textit{Sarsa} algorithm was not designed for the interactive task of human exploration. Reciprocally, this may be induced by experiential factors due to the restricted interaction of our RL agent prototype, \textit{e.g.,} preventing users to undo their last actions. Finally, two participants also complained about the lack of precision of the agent toward the generated sounds. This was induced by the tabular method that we used with the \textit{Sarsa} algorithm, which required to discretize the VST parameter space.

\subsubsection{Design Implications}\label{subsubsec:imply}

Participants jointly expressed the wish to lead agent exploration. They suggested different improvements toward our RL agent prototype:

\begin{itemize}
\item Express richer feedback to the agent (\textit{e.g.}, differentiating \textit{``I like''} from \textit{``I really like''})
\item Control agent path more directly (\textit{e.g.}, commanding the agent to go back to a previous state, or to some new unvisited state in the parameter space)
\item Improve agent algorithm (\textit{e.g.}, acting more precisely on parameters, reacting more directly to feedback)
\item Integrate agent in standard workspace (\textit{e.g.}, directly manipulating knobs at times in lieu of the agent)
\end{itemize}

Interestingly, one participant suggested moving from current sequential workflow (where the agent waits for user feedback to take an action on the environment's state) to an autonomous exploration workflow (where the agent would continuously take actions on the environment's state, based on both accumulated and instantaneous user feedback). Three participants envision that such an improved RL agent could be useful in their practice, potentially allowing for more creative partnerships between users and agents.

\section{Co-Explorer}\label{sec:coax}

Our pilot study led us to the design of a final prototype, called Co-Explorer. We decided to first design new generic interaction modalities with RL agents, based on users' reactions with both parametric interfaces and our initial prototype. We then engineered these interaction modalities, developing a generic deep reinforcement learning algorithm fostering human exploration over learning one optimal behaviour, along with a new specific interface for sound design.

\subsection{Interaction Modalities}\label{subsec:interaction}

\begin{figure}[!h]
  \includegraphics[width=\columnwidth]{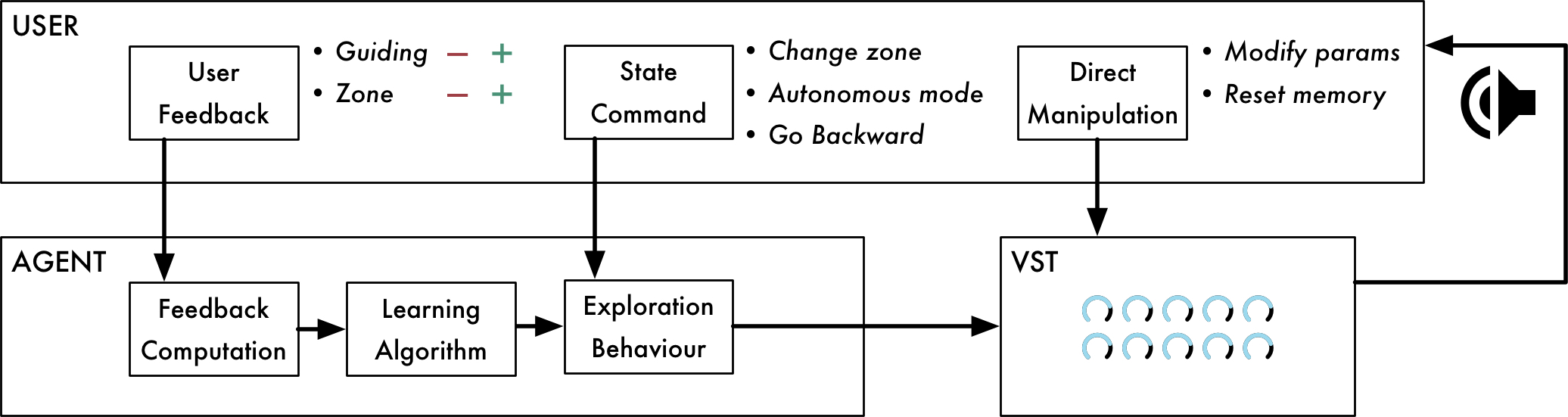}
  \caption{Co-Explorer workflow. Users can have the agent explore parameters autonomously, communicating feedback and state commands to influence agent's actions. Reciprocally, they can directly explore parameters by hand, using a standard parametric interface. Users may be free to switch between these interactions modalities throughout their exploration.}
  \label{fig:workflow}
\end{figure}

Our initial prototype only employed user feedback as its unique interaction modality. This limited our participants, who suggested a variety of new agent controls to support exploration. We translated these suggestions into new interaction modalities that we conceptualized under three generic categories: (1) user feedback, (2) state commands, and (3) direct manipulations (as shown in Fig. \ref{fig:workflow}).

\subsubsection{User Feedback}\label{subsubsec:feedback}

Our design intention is to support deeper user customization of the parameter space, as suggested by our users in the pilot study, who wanted to ``express richer feedback to the agent''. We thus propose to enhance user feedback as defined in our initial prototype, distinguishing between \textit{guiding} and \textit{zone feedback}. Guiding feedback corresponds to users giving binary guidance toward the agent's instantaneous actions in the parameter space. Users can give either positive---\textit{i.e.,} ``keep going in that direction''---or negative guidance feedback---\textit{i.e.,} ``avoid going in that direction''. Zone feedback corresponds to users putting binary preference labels on given state zones in the parameter space. It can either be positive---\textit{i.e.,} ``this zone interests me''---or negative---\textit{i.e.,} ``this zone does not interest me''. Zone feedback would be used for making assertive customization choices in the design space, while guiding feedback would be used for communicating on-the-fly advice to the learning agent. 

\subsubsection{State Commands}\label{subsubsec:command}

Additionally, our design intention is to support an active user understanding of agent trajectory in the parameter space, as suggested by our users in the pilot study, who wanted to ``control agent path more directly''. We propose to define an additional type of interaction modality---we call them ``state commands''. State commands enable direct control of agent exploration in the parameter space, without contributing to its learning. We first allow users to command the agent to \textit{go backward} to some previously-visited state. We also enable users to command the agent to \textit{change zone} in the parameter space, which corresponds to the agent making an abrupt jump to an unexplored parameter configuration. Last but not least, we propose to let users start/stop an \textit{autonomous exploration mode}. Starting autonomous exploration corresponds to letting the agent act continuously on parameters. As such, in autonomous exploration mode, the agent does not have to wait for feedback from the user to take actions. Thus, two different cases arise, at each time step. If the user gives feedback, then the next action is taken on that basis. If the user does not give feedback, then the next action is taken based on past accumulated user feedback. Stopping autonomous exploration corresponds to going back to the sequential workflow implemented in our initial prototype, where the agent waits for user feedback before taking a new action on parameters.

\subsubsection{Direct Manipulation}\label{subsubsec:direct}

Lastly, our design intention is to augment, rather than replace, parametric interfaces with interactive reinforcement learning, as suggested by our users in the pilot study, who wanted to ``integrate agent in standard workspace''. We thus propose to add ``direct manipulations'' to support direct parameter modification through a standard parametric interface. It lets users explore the space on their own by only manipulating parameters without using the agent at all. It can also be used to take the agent to a given point in the parameter space---\textit{i.e.,} ``start exploration from this state''---, or to define by hand certain zones of interest using a zone feedback---\textit{i.e.,} ``this example preset interests me''. Inversely, the parametric interface also allows to visualize agent exploration in real-time by observing how it acts on parameters.

A last, global interaction modality consists in \textit{resetting agent memory}. This enables users to start exploration from scratch by having the agent forget accumulated feedback. Other modalities were considered, such as modifying the agent's speed and precision. Preliminary tests pushed us to decide not to integrate them in the Co-Explorer.

\subsection{Deep Reinforcement Learning}\label{subsec:agent}

As suggested by our users in the pilot study, who wanted to ``improve agent algorithm'', we developed a deep reinforcement learning agent at three intertwined technical levels. Our approach is based on \textit{Deep TAMER} \cite{warnell2017deep} for feedback formalization (Section \ref{subsubsec:feedbackform}), and learning algorithm (Section \ref{subsubsec:algo}). Our original adaptations lie in exploration behaviour (Section \ref{subsubsec:exploration}), and the integration of our interaction modalities in the deep reinforcement learning framework (Section \ref{subsec:modalities}).

\subsubsection{Feedback Formalization}\label{subsubsec:feedbackform}

One challenge consisted in addressing the non-stationarity of user feedback data along their exploration. We implemented \textit{Deep TAMER}, a reinforcement learning algorithm suited for human interaction \cite{warnell2017deep}. \textit{Deep TAMER} leverages a feedback formalization that distinguishes between the environmental reward signal---\textit{i.e.,} named $R$ in the \textit{Sarsa} algorithm of our initial prototype---and the human reinforcement signal---\textit{e.g.,} feedback provided by a human user. This technique, already implemented in the \textit{TAMER} algorithm \cite{knox2009interactively}, was shown to reduce sample complexity over standard reinforcement learning agents, while also allowing human users to teach agents a variety of behaviours.

\textit{Deep TAMER} learns a user's reinforcement function by maximizing direct user feedback. This approach differs from conventional uses of RL, which seek to learn an optimal RL-policy by maximizing cumulative feedback. Yet, optimal RL-policies or fully-trained agents could not be suited to our application, since they make the assumption that users would provide consistent feedback all along the parameter exploration task. Despite being unusual from a RL perspective, we will show that this interactive RL formulation does suit our HCI application.

To deal with potential time delays between reinforcement communicated by the human and state-action paths made by the agent, \textit{Deep TAMER} uses a weighting function $u(t)$ distributing credit over the sequence of lastly-visited state-action pairs. We set $u(t)$ similarly to the \textit{Deep TAMER} implementation, that is, with a uniform distribution for the rewards received between 4 and 0.2 seconds before the most recent reward---\textit{i.e.,} the most recent user feedback, as formulated in Section \ref{subsec:interaction}. We detail the differences between standard RL algorithms and \textit{Deep TAMER} in Appendix A.

\subsubsection{Learning Algorithm}\label{subsubsec:algo}

Another challenge was to tackle learning in high-dimensional parametric spaces that are typical of our use case. \textit{Deep TAMER} employs function approximation \cite{Sutton2011} to generalize user feedback given on a subset of state-action pairs to unvisited state-action pairs. Specifically, a deep neural network is used to learn the best actions to take in a given environment state, by predicting the amount of user feedback it will receive \cite{mnih2015human,warnell2017deep}. The resulting algorithm can learn in high-dimensional state spaces $\mathcal{S}=\{S\}$ and is robust to changes in discretization $a_i$ of the space. For our application in sound design, we engineered the algorithm for $n=10$ parameters. We normalized all parameters and set the agent's precision by discretizing the state space in one hundred levels ($s_i \in \lbrack0,1\rbrack,a_i = 0.01; 0 \leq i \leq n$).

A last challenge was to learn quickly from the small amounts of data provided by users during interaction. \textit{Deep TAMER} uses a replay memory, which consists in storing the received human feedback in a buffer $\mathcal{D}$, and sampling repeatedly from this buffer with replacement \cite{warnell2017deep}. This was shown to improve the learning of the deep neural network in high-dimensional parameter spaces in the relatively short amount of time devoted to human interaction. We set the hyperparameters of the deep neural network by performing a parameter sweep and leading sanity checks with the algorithm; we report them in Appendix B.

\subsubsection{Exploration Behaviour}\label{subsubsec:exploration}

\begin{figure}[h]
  \includegraphics[width=0.65\columnwidth]{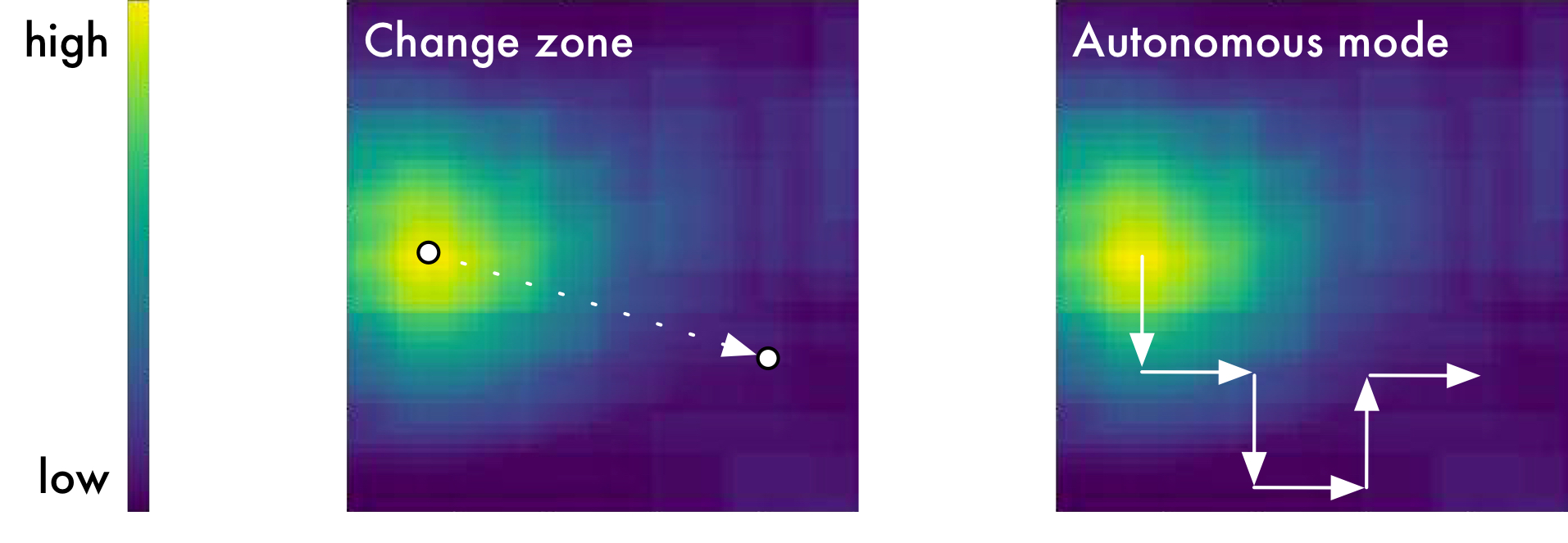}
  \caption{Schematic representations for exploration behaviour. The color scale depicts the density model for a two-dimensional state space. Left: Changing zone has the agent jump to a state with lowest density. Right: Autonomous exploration has the agent take successive actions toward states with lowest density.}
  \label{fig:explore}
\end{figure}

We developed a novel exploration method for autonomous exploration behaviour (see Fig. \ref{fig:explore}). It builds on an intrinsic motivation method, which pushes the agent to ``explore what surprises it'' \cite{bellemare2016unifying}. Specifically, it has the agent direct its exploratory actions toward uncharted parts of the space, rather than simply making random moves---as in the $\varepsilon$-greedy approach implemented in our initial prototype. Our method is based on state visitation density to push the agent to take exploratory actions toward unknown state zones. It builds a density model of the parameter space based on an estimation of state-visitation counts, called the pseudo-count $\hat{V}(S)$, and a total visit pseudo-count $\hat{v}$ using a density model $\hat{p}_{\phi}(S)$. It then adds a reward bonus, $R^+$, to the agent, based on the novelty of the state. We parameterized $\varepsilon$ with an exponential decay in such a way that its initial value would slowly decrease along user exploration. For our application in sound design, agent speed in autonomous exploration mode was set to one action by tenths of a second. We report the hyperparameters set for our exploration method after sanity checks in Appendix B, and detail the density model in Appendix C.

We used tile coding, a specific feature representation extensively used in the reinforcement learning literature to efficiently compute and update the density model $\hat{p}_{\phi}(S)$ in high-dimensional spaces \cite{Sutton2011,watkins1989learning}. To our knowledge, it has not been used for density estimation or in relation with the pseudo-count technique as used here. Tile coding as density estimation techniques was preferred over other techniques such as Gaussian Mixture Models or using Artificial Neural Networks for its low computational cost and ability to scale to higher dimensions. Other exploration methods are based on Thompson sampling \cite{strens2000bayesian}, bootstrapping neural networks for deep exploration \cite{osband2016deep} or by adding parametric noise to network weights \cite{fortunato2017noisy}. Approaches such as Thompson sampling have been used to find an appropriate exploration-exploitation balance but require a prior distribution on the parameter space. Bayesian methods can even be used to compute an optimal exploration-exploitation balance but often require much too great computation resources for the high-dimensional state-action spaces considered in reinforcement learning.

\subsection{Integrating Interaction Modalities In Reinforcement Learning}\label{subsec:modalities}

To fully realize our interaction design, we integrated the modalities defined in Section \ref{subsec:interaction} within the reinforcement learning framework defined in Section \ref{subsec:agent}.

\subsubsection{User Feedback}

\begin{figure}[h]
  \includegraphics[width=0.5\columnwidth]{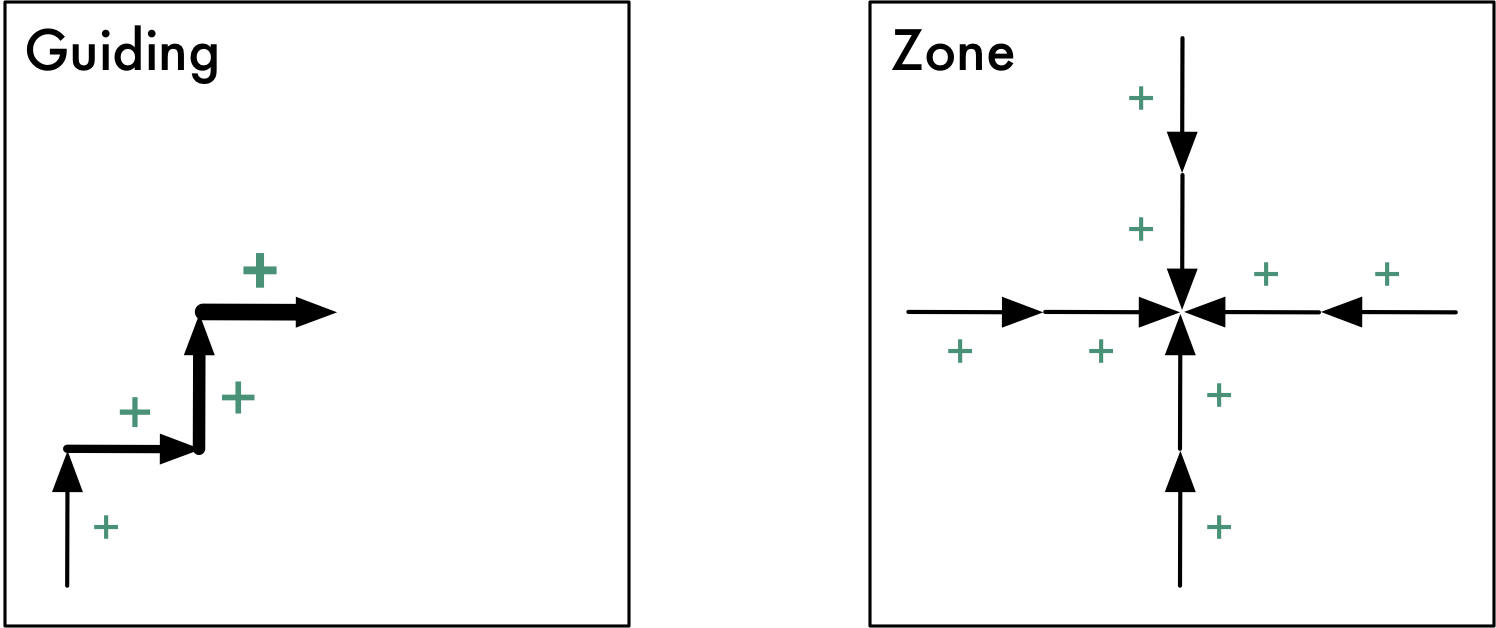}
  \caption{Schematic representations for feedback computation methods. Here, positive feedback is given in some state situated at the center of the square. Left: Guiding feedback is distributed over the $p$ lastly-visited state-action pairs. Right: Zone feedback impacts all state-action pairs potentially leading to the labelled state.}
  \label{fig:feedback}
\end{figure}

We developed generic methods corresponding to user feedback modalities defined in Section \ref{subsubsec:feedback} that we used in the feedback formalization of Section \ref{subsubsec:feedbackform}. For guiding feedback, we assigned user positive or negative feedback value over the $p$ last state-action pairs taken by the agent (see Fig. \ref{fig:feedback}, left), with a decreasing credit given by a Gamma distribution \cite{knox2009interactively}. For zone feedback, we computed all possible state-action pairs leading to the state being labelled and impacted them with positive or negative feedback received (see Fig. \ref{fig:feedback}, right). This enables to build attractive and repulsive zones for the agent in the parameter space. This reward bonus is computed using the density model described in Section \ref{subsubsec:exploration}.

\subsubsection{State Commands}

We developed generic methods corresponding to state commands defined in Section \ref{subsubsec:command} using the exploration behaviour defined in Section \ref{subsubsec:exploration}. Changing zone has the agent randomly sampling the density distribution and jump to the state with lowest density (see Fig. \ref{fig:explore}, left). Autonomous exploration mode has the agent take exploratory actions that lead to the nearest state with lowest density with probability $\varepsilon$ (see Fig. \ref{fig:explore}, right).

\subsubsection{Direct Manipulation}

We integrated direct manipulations as defined in Section \ref{subsubsec:direct} by leveraging the learning algorithm defined in Section \ref{subsubsec:algo}. When parameters are modified by the user, the reinforcement learning agent converts all parameters' numerical values as a state representation, taking advantage of the algorithm's robustness in changes of discretization. As such, direct manipulation is almost continuous in the Co-Explorer, which strongly differs from the coarse-grained, three-level implementation of our initial RL agent prototype. Finally, reseting agent memory has the reinforcement learning algorithm erase all stored user feedback and trajectory, and load a new model.

\subsection{Implementation}

\subsubsection{Agent}

We implemented the Co-Explorer as a Python library\footnote{https://github.com/Ircam-RnD/coexplorer}. It allows to interface the deep reinforcement learning agent to any external input device and output software, using the OSC protocol for message communication \cite{wright2005open}. This was done to enable future applications outside the sound design domain. Each of the features described in Section \ref{subsec:agent} are implemented as parameterized functions, which supports experimentation of interactive reinforcement learning with various parameter values as well as order of function calls. The current version relies on TensorFlow \cite{abadi2016tensorflow} for deep neural network computations. The complete algorithm implementation is described in Appendix D.

\begin{figure}[h]
  \includegraphics[width=\columnwidth]{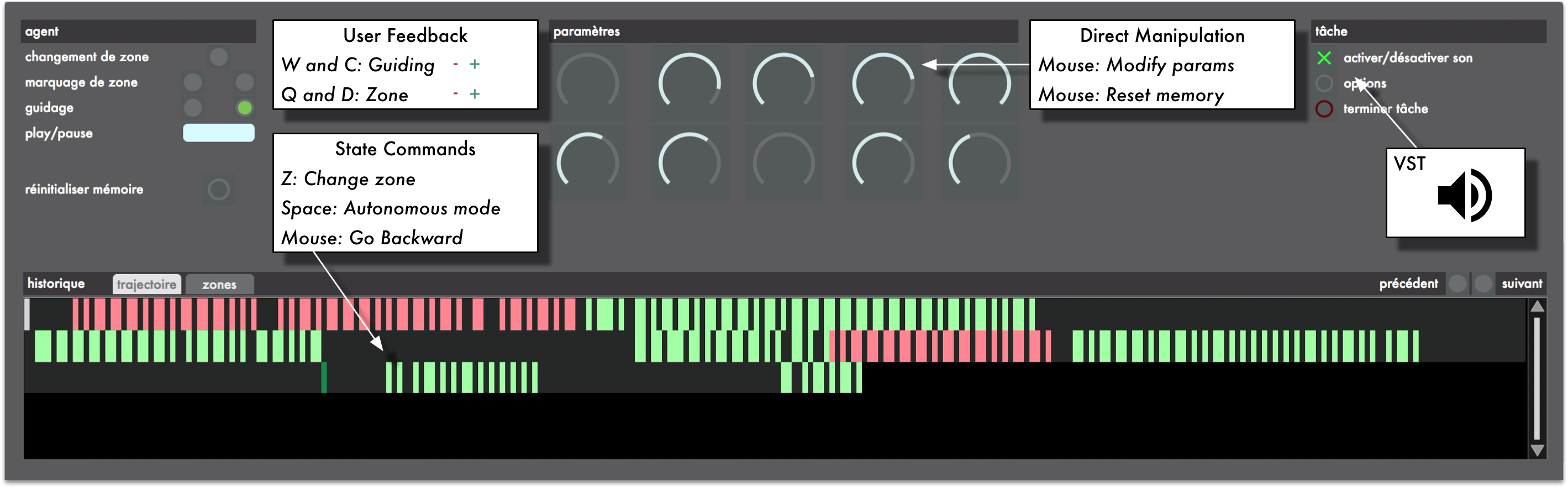}
  \caption{Co-Explorer interface.}
  \label{fig:interface}
\end{figure}

\subsubsection{Interface}

We implemented an interactive interface for our application in sound design (Fig. \ref{fig:interface}), which integrates all interaction modalities defined in Section \ref{subsec:interaction}. It builds on Max/MSP, a visual programming environment for real-time sound synthesis and processing. Standard parametric knobs enable users to directly manipulate parameters, as well as to see the agent act on it in real-time. An interactive history allows users to command the agent to go to a previously-visited state, be they affected by user feedback (red for negative, green for positive) or simply passed through (grey). Keyboard inputs support user feedback communication, as well as state commands that control agent exploration (changing zone, and start/stop autonomous exploration mode). Lastly, a clickable button enables users to reset agent memory.

\section{Evaluation Workshop}

We evaluated the Co-Explorer in a workshop with a total of 12 professional users (5 female, 7 male). The aims of the workshop were to: Evaluate each interaction modality at stake in the Co-Explorer; understand how users may appropriate the agent to support parameter space exploration.

The workshop was divided in two tasks: (1) explore to discover, and (2) explore to create. This structure was intended to test the Co-Explorer in two different creative tasks (described in Section \ref{subsec:discover} and \ref{subsec:create}, respectively). Participants ranged from sound designers, composers, musicians, and artists to music researchers and teachers. They were introduced to the agent's interactive modalities and its internal functioning at the beginning of the workshop. In each part, they were asked to report their observations by filling a browser-based individual journal. Group discussion was carried on at the end of the workshop to let participants exchange views over parameter space exploration. The workshop lasted approximately three hours.

\subsection{Part 1: Explore to Discover}\label{subsec:discover}

\subsubsection{Procedure}

\begin{figure}[h]
  \includegraphics[width=.625\columnwidth]{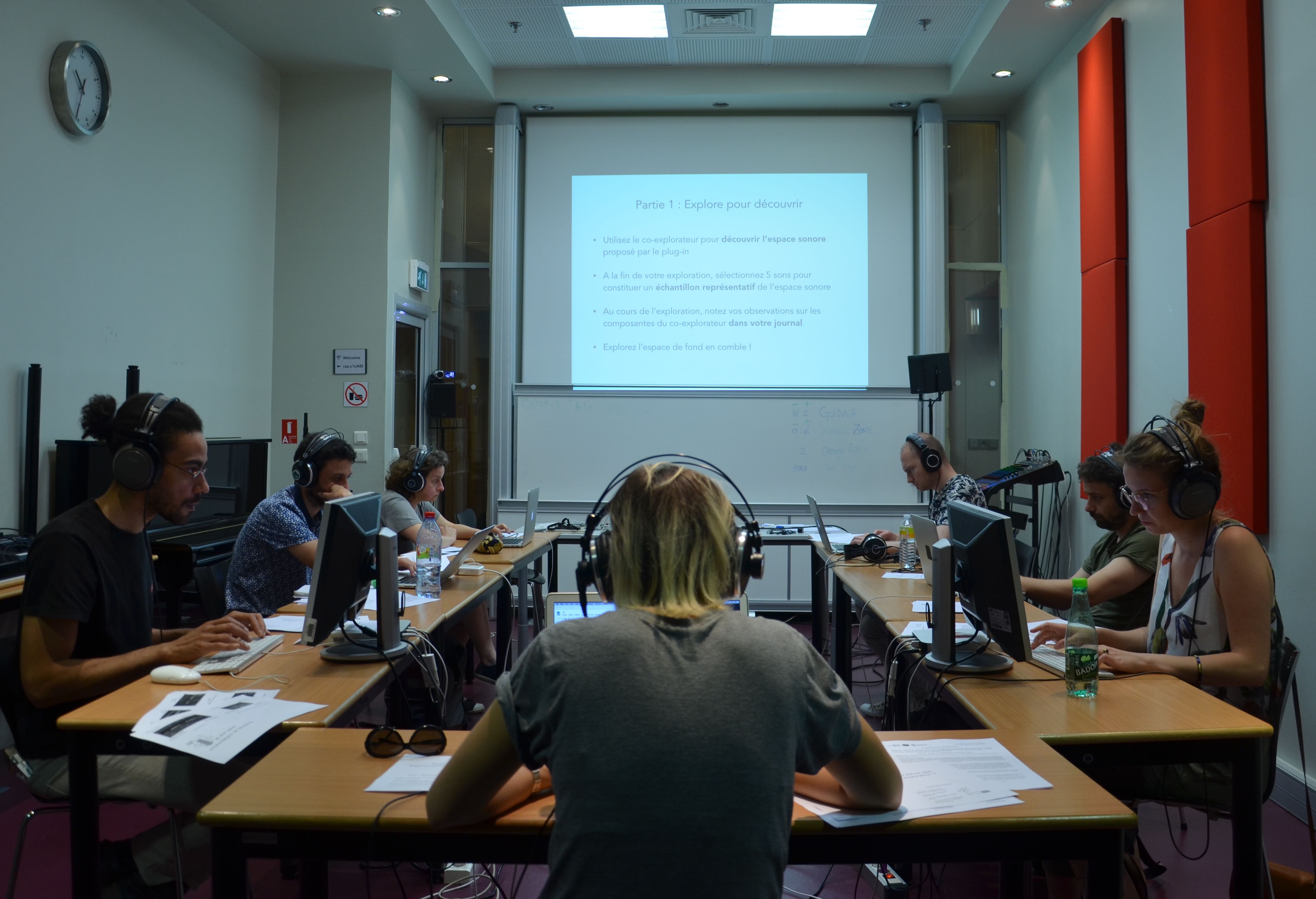}
  \caption{Our participants testing the Co-Explorer in the evaluation workshop.}
  \label{fig:workshop1}
\end{figure}

In the first part of the workshop, participants were presented with one parameter space (see Fig. \ref{fig:workshop1}). They were asked to use the Co-Explorer to explore and discover the sound space at stake. Specifically, we asked them to find and select five presets to constitute a representative sample of the space. We defined the parameter space by selecting ten parameters from a commercial VST. Participants were encouraged to explore the space thoroughly. The task took place after a 10-minute familiarizing session: individual exploration lasted 25 minutes, followed by 5 minutes of sample selection, and 20 minutes of group discussion. Each session started with a fully-untrained agent.

\subsubsection{Analysis}

All participant's actions were logged into a file. These contained timed onsets for user feedback---\textit{i.e.,} binary guiding and zone feedback---, state commands---\textit{i.e.,} backward commands in the history, changing zone commands, and autonomous exploration starting/stopping---, and direct manipulations---\textit{i.e.,} parameter temporal evolutions. We also logged timed onsets for preset selection in relation to the task, but did not include the five presets themselves into our analysis. Our motivation was to focus on the process of exploration in cooperation with the Co-Explorer, rather than on the output of it. We used structured observation to extract information from individual journals and group discussion.

\subsubsection{Results}\label{subsubsec:results}

We first looked at how users employed state commands. Specifically, the autonomous exploration mode, which consisted in letting the agent act cotinuously on parameters on its own, was an important new feature compared to our sequentiam initial RL agent prototype. Participants spent more than half of the task using the Co-Explorer in this mode (total of 13 minutes on average, $\sigma=4.7$). Ten participants used autonomous exploration over several short time slices (average of 50 seconds, $\sigma=25$s), while the two remaining participants used it over one single long period (respectively 9 and 21 minutes). P5 commented about the experience: \textit{``It created beautiful moments during which I really felt that I could anticipate what it was doing. That was when I really understood the collaborative side of artificial intelligence''}. 

The changing zone command, which enabled to jump to an unexplored zone in the parameter space, was judged efficient by all participants to find diverse sounds within the design space. It was used between 14 and 90 times, either to start a new exploration (P1: \textit{``Every time I used it, I found myself in a zone that was sufficiently diametrically opposed to feel that I could explore something relatively new''}), or to rapidly seize the design space in the context of the task (P12: \textit{``I felt it was easy to manage to touch the edges of all opposite textures''}). Interestingly, P2 noticed that the intrinsic motivation method used for agent exploration behaviour \textit{``brought something more than a simple random function that is often very frustrating''}.

We then looked at how users employed feedback. Guiding feedback, enabling guidance toward agent actions, was effectively used in conjunction with autonomous exploration by all participants, balancing positive with negative (55\% positive on average, $\sigma=17\%$). Participants gave various amounts of guiding feedback (between 54 and 1489 times). These strategies were reflected by different reactions toward the Co-Explorer. For example, one participant was uncertain in controlling the agent through feedback: \textit{``if the agent goes in the right direction, I feel like I should take time to see where it goes''}, he commented. On the contrary, P1 was radical in his controlling the agent, stating that he is \textit{``just looking for another direction''}, and that he uses feedback \textit{``without any value judgement''}. This reflects the results described in Section \ref{subsubsec:reactions} using our initial RL agent prototype.

Zone feedback, enabling customization of the space with binary state labels, was mostly given as positive by participants (72\%, $\sigma=18\%$). Two participants found the concept of negative zones to be counter-intuitive. \textit{``I was a bit afraid that if I label a zone as negative, I could not explore a certain part of the space''}, P8 coined. This goes in line with previous results on applying interactive reinforcement learning in the field of robotics \cite{thomaz2008teachable}. All participants agreed on the practicality of combining positive zone feedback with backward state commands in the history to complete the task. \textit{``I labeled a whole bunch of presets that I found interesting [...] to after go back in the trajectory to compare how different the sounds were, and after continue going in other zones. I found it very practical''}, P8 reported. Overall, zone feedback was less times used than guiding feedback (between 10 and 233 times).

Finally, direct manipulation was deemed efficient by participants in certain zones of the design space. \textit{``When I manage to hear that there is too much of something, it is quicker to parametrize sound by hand than to wait for the agent to find it itself, or to learn to detect it''}, P4 analyzed. P10 used them after giving a backward state command, saying she \textit{``found it great in cases where one is frustrated not to manage to guide the agent''}. P11 added that she directly manipulate parameters to \textit{``adjust the little sounds that [she] selected''}. P1 suggested that watching parameters move as the agent manipulates them could help learn the interface: \textit{``From a pedagogical point of view, [the agent] allows to access to the parameters' functioning and to the interaction between these parameters more easily [than without]''}. This supports the fact that machine learning visualizations may be primordial in human-centred applications to enable interpretability of models \cite{amershi2014power}.

\subsubsection{Relevance to Task}

Three participants wished that the Co-Explorer reacted more quickly to feedback in relation to the task: \textit{``I would really like to feel the contribution of the agent, but I couldn't''}, P12 said. Also, P3 highlighted the difficulties to give evaluative feedback in the considered task: \textit{``without a context, I find it hard''}, he analysed. Despite this, all participants wished to spend more time teaching the Co-Explorer, by carefully customizing the parameter space with user feedback. For example, five participants wanted to slow the speed of the agent during autonomous exploration to be able to give more precise guidance feedback. Also, three participants wanted to express sound-related feedback:  \textit{``There, I am going to guide you about the color of the spectrum. [...] There, I'm going to guide you about, I don't know, the harmonic richness of the sound, that kind of stuff...''}, P4 imagined.

\subsection{Part 2: Explore to Create}\label{subsec:create}

\subsubsection{Procedure}

\begin{figure}[!h]
  \includegraphics[width=0.5\columnwidth]{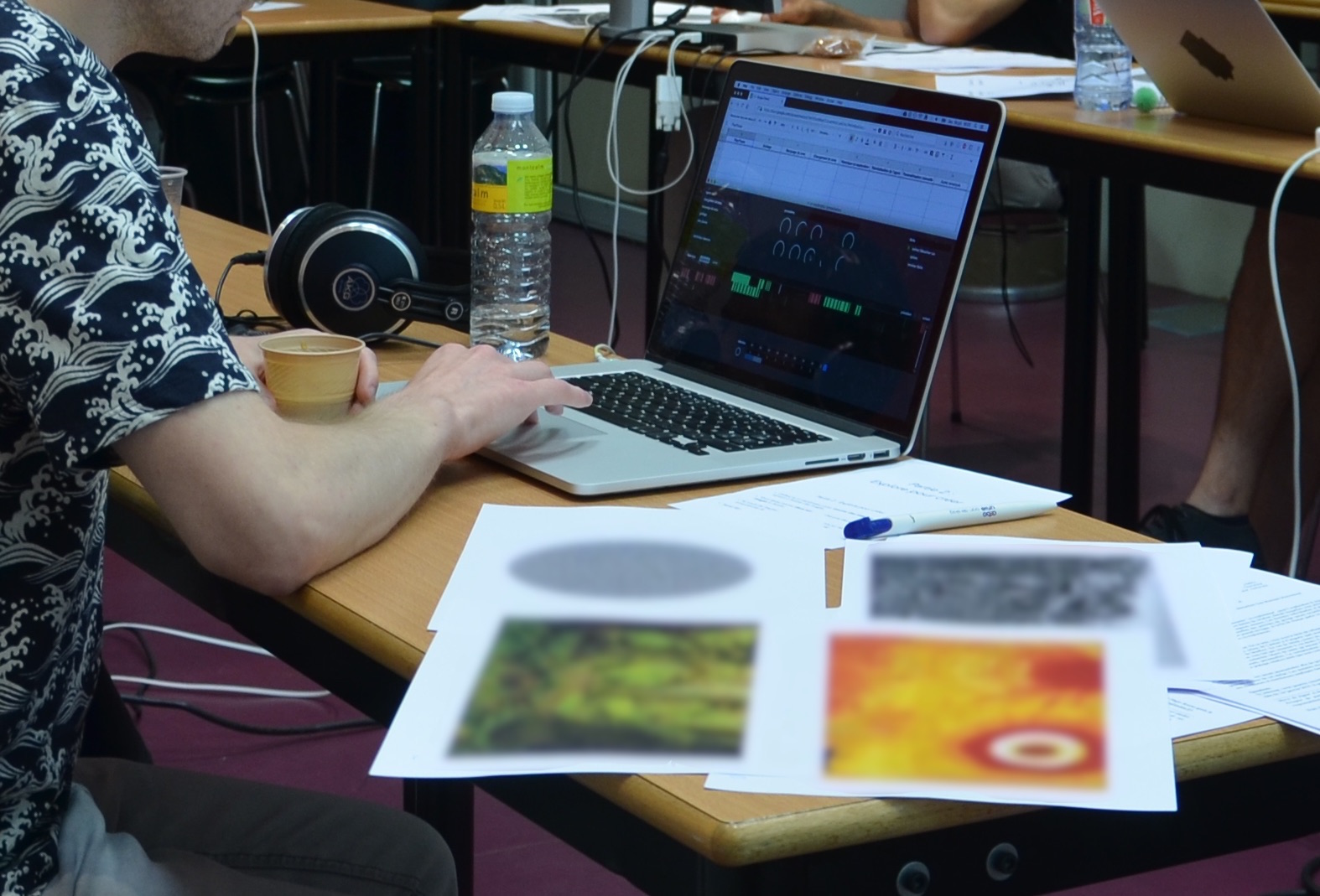}
  \caption{The four pictures framing the creation task of the workshop.}
  \label{fig:workshop2}
\end{figure}

In the second part of the workshop, participants were presented with four pictures (Fig. \ref{fig:workshop2}) created by renowned artists and photographers. For each of these four pictures, they were asked to explore and create two sounds that subjectively depict the atmosphere of the picture. In this part, we encouraged participants to appropriate interaction with the Co-Explorer and feel free to work as they see fit. We used a new sound design space for this second part, which we designed by selecting another ten parameters from the same commercial VST than in Part 1. Individual exploration and sound selection lasted 30 minutes, followed by 20 minutes of group discussion and 10 minutes of closing discussion. The session started with a fully-untrained agent.

\subsubsection{Analysis}

All participant actions were logged into a file, along with timed parameter presets selected for the four pictures. Again, we focused our analysis on the process of exploration rather than on the output of it. Specifically, for this open-ended, creative task, we did not aim at analysing how each agent interaction modality individually relates to a specific user intention. Rather, we were interested in observing how users may appropriate the mixed-initiative workflow at stake in the Co-Explorer.

We used Principal Component Analysis (PCA \cite{jolliffe2011principal}), a dimensionality reduction method, to visualize how users switched parameter manipulation with agents. We first concatenated all participants' parameter evolution data as an $n$-dimensional vector to compute the two first principal components. We then projected each participant data onto these two components to support analysis of each user trajectory on a common basis. By doing this, relatively distant points would correspond to abrupt changes made in parameters (\textit{i.e.}, to moments when the user takes the lead on exploration). Continuous lines would correspond to step-by-step changes in parameters (\textit{i.e.}, to moments when the Co-Explorer explores autonomously). PCA had a stronger effect in the second part of our workshop. We interpret this as a support to the two-part structure that we designed for the workshop, and thus did not include analysis of the first part. Finally, we used structured observation to extract information from individual journals and group discussion.

\subsubsection{Exploration Strategies}\label{subsubsec:coexstrat}

\begin{figure}[!h]
  \includegraphics[width=.625\columnwidth]{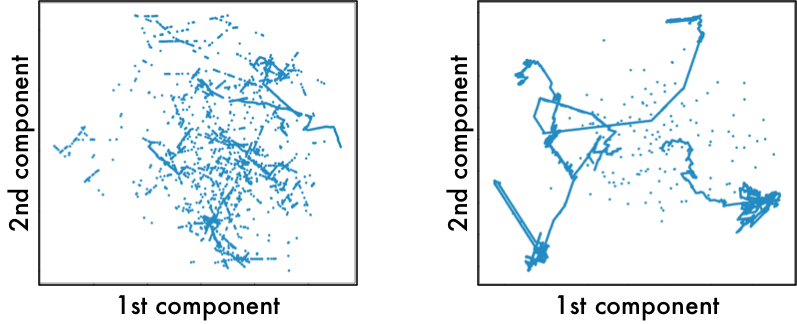}
  \caption{Two types of co-exploration partnerships shown in PCA visualizations of parameter evolution: User-as-leader (P9, left) and agent-as-leader (P7, right). Relatively distant points correspond to abrupt changes made in parameters (\textit{i.e.}, to moments when the user takes the lead). Continuous lines correspond to step-by-step changes in parameters (\textit{i.e.}, to moments when the Co-Explorer takes the lead).}
  \label{fig:paths}
\end{figure}

All participants globally expressed more ease interacting with the Co-Explorer in this second task. \textit{``I felt that the agent was more adapted to such a creative, subjective... also more abstract task, where you have to illustrate. It's less quantitative than the first task''}, P9 analysed. User feedback was also reported to be more intuitive when related to a creative goal: \textit{``all parameters took their sense in a creative context. [...] I quickly found a way to work with it that was very efficient and enjoyable''}, P5 commented. Figure \ref{fig:paths} illustrates the PCA for two different users.

Qualitative analysis of PCAs let us conceptualize a continuum of partnerships between our participants and the Co-Explorer. These could be placed anywhere between the two next endpoints:
\begin{itemize}
\item \textbf{User-as-leader}: This workflow involves users first building a customized map of the design space, then generating variations over these custom presets. In terms of interaction modalities, this consists in first iteratively using changing zone and positive zone feedback to store custom presets, then either using direct manipulation or short autonomous explorations to generate variations of these presets.
\item \textbf{Agent-as-leader}: This workflow involves users letting the Co-Explorer lead most of parameter manipulation. In terms of interaction modalities, this consists in using autonomous exploration mode combined with guiding feedback over long periods of time, occasionally using changing zone or direct manipulation to choose a start point for the Co-Explorer to lead autonomous exploration.
\end{itemize}

Our interpretation is as follows. User-as-leader partnership may correspond to user profiles that approach creative work as a goal-oriented task, where efficacy and control are crucial (P10: \textit{``I am accustomed... Where I work, if you prefer, we have to get as quick as possible to the thing that works the best, say, and I cannot spend so much time listening to the agent wandering around''}). Reciprocally, agent-as-leader partnership may correspond to user profiles that approach creative work as an open-ended task, where serendipity is essential for inspiration (P5: \textit{``I did not try to look for the sound that would work the best. I rather let myself be pushed around, even a bit more than in my own practice''}). Some participants did not stabilize into one single partnership, but rather enjoyed the flexibility of the agent. \textit{``It was quite fun to be able to let the agent explore, then stop, modulate a bit some parameters by hand, let it go and guide it again, changing zones too, then going back in the history... Globally, I had the impression of shaping, somewhat... I found it interesting''}, P11 coined.

Agent memory was handled with relevance to various creative processes toward the pictures. Seven participants disposed all four pictures in front of them (P7: \textit{``to always have them in mind. Then, depending on the agent's exploration, I told myself `hey, this sound might correspond to this picture'''}). Three participants focused on one picture at a time, \textit{``without looking at the others''}. Four participants never reset the memory (P11: \textit{``my question was, rather, in this given sonic landscape, how can I handle these four pictures, and reciprocally''}), and three participants reset agent memory for each of the different atmospheres shared by the pictures. Overall, participants benefited from partnering with the Co-Explorer in parameter space exploration: \textit{``It's a mix of both. I easily managed to project a sound on the picture at first glance, then depending on what was proposed, it gave birth to many ideas''}, one participant said.

\subsubsection{Toward Real-World Usages}

All participants were able to describe additional features for the Co-Explorer to be usable in their real-world professional work environments---examples are, among others, connection to other sound spaces, memory transfer from one space to another, multiple agent memory management, or data exportation. They also anticipated creative uses for which the Co-Explorer were not initially designed. Half of the participants were enthusiastic about exploiting the temporal trajectories as actual artifacts of their creation (P6: \textit{``What I would find super interesting is to be able to select the sequences corresponding to certain parameter evolution, or playing modes. [...] It would be super great to select and memorize this evolution, rather than just a small sonic fragment''}). Finally, two participants further imagined the Co-Explorer to be used as musical colleagues---either as improvisers with which one could \textit{``play with both hands''} (P2), or as \textit{``piece generators''} (P6) themselves.

\section{Discussion}\label{sec:discussion}

Our process of research, design, and development led to contributions at three different levels: (1) conceptual insight on human exploration; (2) technical insight on reinforcement learning; and (3) joint conceptual and technical design guidelines on machine learning for human creativity.

\subsection{Conceptual Insight}

\subsubsection{From Exploration to Co-Exploration}

Our work with interactive reinforcement learning allowed for observing and characterizing user approaches to parameter space exploration, and supported it. While manipulating unlabelled parametric knobs of sound synthesizers, participants alternated between an \textit{analytical} approach---attempting to understand the individual role of each parameter---and a \textit{spontaneous} approach that could lead to combinations in the parameter space that might not be guessed with the analytical approach. While interacting with a reinforcement learning agent, participants tended to alternate the lead in new types of mixed-initiative workflows \cite{horvitz1999principles} that we propose to call \textit{co-exploration} workflows. \textit{User-as-leader} workflow was used for gaining control over each parameter of the design space. \textit{Agent-as-leader} workflow allowed to relax users' control and provoke discoveries through the specific paths autonomously taken by the agent in the parameter space. 
Importantly, the benefit of interactive reinforcement learning for co-exploring sound spaces was dependent on the task. We found that this co-exploration workflow were more relevant to human exploration tasks that have a focus on creativity, such as in our workshop's second task, rather than discovery. Therefore, we believe that this workflow is well-suited in cases where exploration is somehow holistic (as in the creative task) rather than analytic (as in the discovery task where the goal is to understand the sound space to find new sounds). In a complementary work, described in Appendix E, we were able to validate this hypothesis, by proving that guiding the RL-agent better supports user creativity than simply using a standard parametric interface.

\subsubsection{Methodology}

Our user-centered design approach to interactive reinforcement learning and exploration allowed us to rapidly evaluate flexible interaction designs without focusing on usability. This process let us discover innovative machine learning uses that we may not have anticipated if we had started our study with an engineering phase. The simple, flexible, and adaptable designs tested in our first pilot study (parametric vs. RL) could in this sense be thought as technology probes \cite{hutchinson2003technology}. Working with professional users of different background and practices---from creative coders to artists less versed in technology---was crucial to include diverse user feedback in the design process. Our results support this, as many user styles were supported by the Co-Explorer. That said, user-driven design arguably conveys inherent biases of users. This is particularly true when promoting AI in interactive technology \cite{amershi2019guidelines,caramiaux2019ai}. As a matter of fact, alongside a general enthusiasm, we did observe a certain ease among our professional users for expressing tough critiques, at times being skeptical on using AI, especially when the perception of the algorithm choice would contradict their spontaneous choice. Yet, the two professional users that took part to both our pilot study and workshop found the use of AI as welcome, testifying of its improvement along the development process.

\subsubsection{Evaluation}

Lastly, evaluation of reinforcement learning tools for creativity remains to be investigated more deeply. While our qualitative approach allowed us to harvest thoughtful user feedback on our prototypes' interaction modalities, it is still hard to account for direct links between agent computations and user creative goals. Using questionnaire methods, such as the Creativity Support Index \cite{cherry2014quantifying}, may enable to measure different dimensions of human creativity in relation to different algorithm implementations. As a first step toward this direction, we report a preliminary summary that maps some of participants' quotes in our evaluation workshop to Creativity Support Index dimensions in Appendix F. Also, focusing on a specific user category could also allow more precise evaluation in relationship to a situated set of creative practices and uses. Alternatively, one could aim at developing new reinforcement learning criteria that extends standard quantitative measures---such as convergence or learning time \cite{Sutton2011}---to the qualitative case of human exploration. Research on interactive supervised learning has shown that criteria usually employed in the field of Machine Learning may not be adapted to users leading creative work \cite{Fiebrink2011}. We believe that both HCI and ML approaches may be required and combined to produce sound scientific knowledge on creativity support evaluation.

\subsection{Technical Insight}

\subsubsection{Computational Framework}

Our two working prototypes confirmed that interactive reinforcement learning may stand as a generic technical framework for parameter space exploration. The computational framework that we proposed in Section \ref{subsubsec:proto}, leveraging states, actions, and rewards, strongly characterized the mixed-initiative co-exploration workflows observed in Section \ref{subsec:create}---\textit{e.g.,} making small steps and continuous trajectories in the parameter space. Other interactive behaviours could have been implemented---\textit{e.g.,} allowing the agent to act on many parameters in only one action, or using different $a_i$ values for different action sizes---to allow for more diverse mixed-initiative behaviours. Alternatively, we envision that domain-specific representations may be a promising approach for extending co-exploration. In the case of sound design, one could engineer high-level state features based on audio descriptors \cite{schwarz2009sound} instead of using raw parameters. This could allow RL agents to learn state-action representations that would be independent from the parameter space explored---potentially allowing memory transfer from one parameter state space to another. This could also enable agent adaptation of action speed and precision based on perceptual features of the parameter space---potentially avoiding abrupt jumps in sound spaces.

\subsubsection{Learning Algorithm}

Reinforcement learning algorithmic functioning, enabling agents to learn actions over states, was of interest for our users, who were enthusiastic in teaching an artificial agent actions by feedback. Our deep reinforcement learning agent is a novel contribution to HCI research compared to multi-armed bandits (which explore actions over one unique state \cite{lomas2016interface}), contextual bandits (which explore in lower-dimensional state spaces \cite{koch2019may}), and bayesian optimization (which explores states at implicit scales \cite{shahriari2016taking}). We purposely implemented heterogeneous ways of teaching with feedback based on our observations of users' approaches to parameter space exploration, which extends previous implementations such as those in the Drawing Apprentice \cite{davis2016empirically}. We also decided to have the agent maximize direct user feedback (for which \textit{Deep TAMER} was adapted \cite{warnell2017deep}, as opposed to \textit{Sarsa} \cite{Sutton2011}), rather than to optimize one general RL policy. Indeed, our observations in the pilot study suggested that exploring users may not generate one goal-oriented feedback signal, but may rather have several sub-optimal goals. They may also make feedback mistakes, act socially toward agents, or even try to trigger surprising agent behaviours over time. In this paper, we focused on qualitative evaluation of the learned policies to provide a proof of interest of interactive reinforcement learning from an HCI perspective. At time of writing, we know of no way to provide quantitative evaluation of the interactive learning of such policies with an exploring user from a machine learning perspective. We believe that such evaluation methods do constitute a matter of research beyond the scope of this paper, which is currently an emerging topic in RL \cite{he2020learning}. Beyond agent customisation, future research may address agent generalisation to other sessions or users, for example having users start a session with a partially-trained agent instead of fully-untrained.

\subsubsection{Exploration Behaviours}

The exploration behaviours of reinforcement learning agents were shown promising for fostering creativity in our users. Both $\varepsilon$-greedy and intrinsic motivation method were adapted to the interactive case of a user leading exploration. One of our users felt that intrinsic motivation had agents behave better than random. In a complementary work \cite{scurto2018perceiving}, we confirmed that users perceived the difference between a random agent and an interactively-learning RL agent. Interestingly, what they perceive may be more related to the agent's global effect in exploring the parameter space, rather than the difference between various implementations of agent exploration. Future work may investigate how user perception of agent exploration may relate to specific implementations of exploration methods. Complementary to such an approach, future work may study co-exploration partnerships in real-world applications to inquire co-adaptation between users and agents over longer periods of time \cite{mackay1990users}. On the one hand, users could be expected to learn to provide better feedback to RL agents to fulfill their creative goals---as it was shown in interactive approaches to supervised learning \cite{Fiebrink2011}. On the other hand, agents could be expected to act more in line with users by exploiting larger amounts of accumulated feedback data---as it is typical with interactive reinforcement learning agents \cite{Sutton2011}. A more pragmatic option would be to give users full control over agent epsilon values---\textit{e.g.,} using an interactive slider \cite{koch2019may}---to improve partnership in this sense.

\subsection{Guidelines for Designing With Machine Learning in Creative Applications}

Based on our work with reinforcement learning, we identified a set of design challenges for leading joint conceptual and technical development of other machine learning frameworks for creative HCI applications. We purposely put back quotes from our participants in this section to inspire readers with insights on AI from users outside our design team.

\subsubsection{Engage Users with Machine Learning}

The Co-Explorer enabled users to fully engage with reinforcement learning computational framework. Users could explore as many states, provide as much feedback, and generate as many agent actions as they wanted to. They also had access to agent memory, be it by navigating in the interactive history, or by reseting the learned behaviour. In this sense, they had full control over the algorithmic learning process of the agent. This is well articulated by a participant, whose quote can be reported here: \textit{``I did not feel as being an adversary to, or manipulated, by the system. A situation that can happen with certain audio software that currently use machine learning, where it is clear that one tries to put you on a given path, which I find frustrating---but this was not the case here''}.

These observations suggest that user engagement at different levels of machine learning processes may be essential to create partnering flows \cite{pachet2013reflexive}. That is, users should be provided with interactive controls and simple information on learning to actively direct co-creation. This goes in line with previous works studying user interaction with supervised learning in creative tasks \cite{amershi2014power}, which showed how users can build better partnerships by spending time engaging with algorithms \cite{Fiebrink2011}. Careful interaction design must be considered to balance full automation with full user control and aim at creating flow states among people \cite{csikszentmihalyi1997flow}. Aiming at such user engagement may also constitute a design opportunity to demystify AI systems, notably by having users learn from experience how algorithms work with data \cite{fiebrink2019machine}.

\subsubsection{Foster Diverse Creative Processes}

Our work showed that the Co-Explorer supported a wide diversity of creative user processes. Users could get involved in open-ended, agent-led exploration, or decide to focus on precise, user-led parameter modification. Importantly, none of these partnerships were clearly conceptualized at the beginning of our development process. Our main focus was to build a reinforcement learning agent able to learn from user feedback and to be easily controllable by users. In this sense, the Co-Explorer was jointly designed and engineered to ensure a dynamic human process rather than a static media outcome. As a matter of fact, we report one participant's own reflection, which we believe illustrate our point: \textit{``What am I actually sampling [from the parameter space]? Is is some kind of climate that is going to direct my creation afterwards? [...] Or am I already creating?''}.

This suggests that supporting the process of user appropriation may be crucial for building creative AI partnerships. Many creative tools based on machine learning often focus on engineering one model to ensure high performance for a given task. While these tools may be useful for creative tasks that have a focus on high productivity, it is arguable whether they may be suited to creative work that has a focus on exploration as a way to build expression. For the latter case, creative AI development should not focus on one given user task, but should rather focus on providing users with a dynamic space for expression allowing many styles of creation \cite{resnick2005design}. The massive training datasets, which are usually employed in the Machine Learning community to build computational creativity tools, may also convey representational and historical biases among end users \cite{suresh2019framework}. Interactive approaches to machine learning directly address this issue by allowing users to intervene in real-time in the learning process \cite{fiebrink2016machine}.

\subsubsection{Steer Users Outside Comfort Zones}

The Co-Explorer actively exposed the exploration behaviour of reinforcement learning to users. This goes in opposition with standard uses of these algorithms \cite{brockman2016openai}, and may provoke moments where agents behaviours may not align with users creative drive \cite{crandall2018cooperating}. Yet, it managed to build \textit{``playful''} and  \textit{``funny''} partnerships that led some users to reconsider their approach to creativity, as one participant confessed: \textit{``At times, the agent forced me to try and hear sounds that I liked less---but at least, this allowed me to visit unusual spaces and imagine new possibilities. This, as a process that I barely perform in my own creative practice, eventually appeared as appealing to me''}.

This suggests that AI may be used beyond customisation aspects to steer users outside their comfort zones in a positive way. That is, designers should exploit non-optimal algorithmic behaviours in machine learning methods to surprise, obstruct, or even challenge users inside their creative process. Data-driven user adaptation may be taken from an opposite side to inspire users from radical opposition and avoid hyper-personalization \cite{andersen2016conversations}. Such an anti-solutionist \cite{blythe2016anti} approach to machine learning may encourage innovative developments that fundamentally reconsider the underlying notion of universal performance commonly at stake in the field of Machine Learning and arguably not adapted to the human users studied in the field of Human-Computer Interaction. It may also allow the building of imperfect AI colleagues, in opposion to ``heroic'' AI colleagues \cite{d2015heroic}: being impressed by the creative qualities of an abstract artificial entity may not be the best alternative to help people develop as creative thinkers \cite{resnick2007all}. The Co-Explorer fairly leans toward such an unconventional design approach, which, in default of fitting every user, surely forms one of its distinctive characteristics.

Several machine learning frameworks remains to be investigated under the light of these human-centred challenges. Evolutionary computation methods \cite{fogel2006evolutionary} may be fertile ground for supporting user exploration and automated refinement of example designs. Active learning methods \cite{settles2010active} may enable communication flows between agents and users that go beyond positive or negative feedback. Dimensionality reduction methods for interactive visualization \cite{maaten2008visualizing} may improve intelligibility of agent actions in large parameter spaces and allow for more trustable partnerships. Ultimately, combining reinforcement learning with supervised learning could offer users with the best of both worlds by supporting both example and feedback inputs. Inverse reinforcement learning \cite{abbeel2004apprenticeship} may stand as a technical framework supporting example input projection and transformation into reward functions in a parameter space.

\section{Conclusion}

In this paper we presented the design of a deep reinforcement learning agent for human parameter space exploration. We worked in close relationship with professional creatives in the field of sound design and led two design iterations during our research process. A first pilot study let us observe users interacting with standard parametric interfaces, as well as with an initial interactive reinforcement learning prototype. The gathered user feedback informed the design of the Co-Explorer, our fully-functioning prototype, for which we led joint design and engineering for the specific task of parameter space exploration. A final workshop allowed us to observe a wide range of partnerships between users and agents, in tasks requiring both quantitative, media-related sampling and qualitative, creative insight.

Our results raised contributions at different levels of research, development, and design. We defined properties of user approaches to parameter space exploration within standard parametric interfaces, as well as to what we called parameter space co-exploration---exploring in cooperation with a reinforcement learning agent. We adapted a deep reinforcement learning algorithm to the specific case of parameter space exploration, developing specific computational methods for user feedback input in high-dimensional spaces, as well as a new algorithm for agent exploration based on intrinsic motivation. We raised general design challenges for guiding the building of new human-AI partnerships, encouraging interdisciplinary research collaborations \cite{pedersen2018twenty} that value human creativity over machine learning performance. We look forward to collaborating with researchers, developers, designers, artists, and users from other domains to take up the societal challenge of designing partnering AI tools that nurture human creativity.

\begin{acks}
We are grateful to our participants for their precious time and feedback. We thank Benjamin Matuszewski, Jean-Philippe Lambert, and Ad\`{e}le P\'{e}cout for their support in designing the studies. This research was partly supported by the ELEMENT project (ANR-18-CE33-0002) from the French National Research Agency.
\end{acks}

\bibliographystyle{ACM-Reference-Format}

\newpage

\label{ap:algorithm}

\section*{Appendix A}

The TAMER \cite{knox2009interactively} and Deep TAMER \cite{warnell2017deep} algorithms can be seen as value-based algorithms. They have been applied in settings that allow to quickly learn a policy on episodic tasks (small game environments or physical models) and aim to maximise direct human reward. This opposed to the traditional RL training objective to maximise the discounted sum of future rewards. These algorithms learn the human reward function $R$ using an artificial neural network and construct a policy from $R$ taking greedy actions. In addition, to accommodate sparse and delayed rewards from larger user response times, the algorithms include a weighting function $u(t)$ to past state trajectories and a replay memory in the case of Deep TAMER. Specifically, while traditional RL algorithms aim to optimise the Mean-Square Error (MSE) loss
\begin{equation}
MSE = \bigg[R_{t+1} + \gamma q(S_{t+1},A_{t+1},\mathbf{w}_t) - q(S_t,A_t,\mathbf{w}_t)\bigg]^2\text{,}
\end{equation}
with $R_t$ the reward at time $t$, $\gamma$ the discount rate, and $q(S_{t},A_{t},\mathbf{w}_t)$ the computed state-action value function with parameters $\mathbf{w}$, (Deep) TAMER aims to optimise
\begin{equation} \label{eq:8}
MSE = u(t) \bigg[R_t - \hat{R}(S_t,A_t)\bigg]^2
\end{equation}
with $R_t$ and $u(t)$ respectively the user-provided feedback and weighting function at time $t$, and $\hat{R}(S_t,A_t)$ the average reward.

\section*{Appendix B}

\begin{table}[!h]
\begin{tabular}{|l|l|}
\hline
\multicolumn{1}{|c|}{\textbf{Deep neural network} \cite{warnell2017deep}} & \multicolumn{1}{c|}{\textbf{Agent} \cite{warnell2017deep}} \\ \hline
\# hidden layers $= 2$ & state dim. $n = 10$ \\ \hline
\# units per layer $= 100$ & $s_i \in [0, 1], 0 \leq i \leq n$ \\ \hline
batch size $= 32$ & $a_i = 0.01, 0 \leq i \leq n$ \\ \hline
learning rate $\alpha = 0.002$ & reward value $|R| = 1$ \\ \hline
replay memory $ \mathcal{D} = 700$ & reward length $= 10$ \\ \hline
\end{tabular}
\begin{tabular}{|l|l|}
\hline
\multicolumn{1}{|c|}{\textbf{Exploration} \cite{warnell2017deep,bellemare2016unifying}} & \multicolumn{1}{c|}{\textbf{Density $\hat{p}_{\phi'}()$}} \\ \hline
$\varepsilon$ decay $= 2000$ & \# tiles $= 64$ \\ \hline
$\varepsilon$ start $= 0.1$ & tile size $= 0.4$ \\ \hline
$\varepsilon$ end $= 0.0$ & $C = 0.01$ \cite{bellemare2016unifying} \\ \hline
action freq. $= 10$ Hz & $\beta = 1$ \cite{bellemare2016unifying} \\ \hline
\end{tabular}
\caption{Hyper-parameters for Deep TAMER (left) and the Co-Explorer exploration behaviour (right).}
\end{table}

\vspace{-0.5cm}
\section*{Appendix C}

The exploration behaviour implemented in the Co-Explorer is based on the notion of optimistic exploration (the assumption that the unknown is good for the agent) and the addition of an `exploration bonus' to the reward. As shown in \cite{bellemare2016unifying}, this exploration bonus can be based on an estimation of state-visitation counts called the pseudo-count $\hat{V}(S)$ and a total visit pseudo-count $\hat{v}$ using a density model $\hat{p}_\phi(S)$. Formally, one calculates the total reward with bonus $R_t^+$ as
\begin{equation}\label{eq:11}
R_t^+ = R_t + \beta \sqrt{\frac{1}{\hat{V}(S_t) + C}}
\end{equation}
with
\begin{equation}
\hat{V}(S_t) = \hat{v}\hat{p}_\phi(S_t)
\end{equation}
\begin{equation}
\hat{v} = \frac{1 - \hat{p}_{\phi'}(S_t)}{\hat{p}_{\phi'}(S_t)-\hat{p}_\phi(S_t)}\hat{p}_\phi(S_t)
\end{equation}
with $\beta$ and $C$ pre-defined constants. We used tile-coding to estimate a density model $\hat{p}_\phi(S)$ over the high-dimensional state spaces considered in our work.

\newpage

\section*{Appendix D}

\begin{algorithm}[H]
\SetAlgoLined
\textbf{Input}: reward function $\hat{R}(S,A,\mathbf{w})$, policy $\pi()$ as $\varepsilon-$greedy with exponential decay, reward distribution function $Env\_dist(R)=R*exp(-t), 0 \leq t \leq R_{length}$ \; 
\textbf{Initialise}: weights $\mathbf{w} = \mathbf{0}$, average-reward $\hat{R} = 0$, $s_i(t=0) = 0.5$ for $0 \leq i \leq n$ and $A(t=0) = \pi(S(t=0))$, $x_j = 0, 0 \leq j \leq R_{length}$ \;
 
 \While (\tcp*[f]{Start autonomous exploration mode}) {running} {
  Take action $A_t$ and observe next state $S_{t+1}$ \;
  Select new action $A_{t+1} \sim \pi(\cdot | S_{t+1})$ \;
  Store $(S_{t+1},A_{t+1}, 0)$ in reward length vector $x$ ($R_{t+1}$ stored as 0)\;
  Update density model $\hat{p}()$ \;
  Observe reward as $R_{t+1}$ \;
  $S_t \leftarrow S_{t+1}$ \;
  $A_t \leftarrow A_{t+1}$ \;
  
  \uIf (\tcp*[f]{Train on user feedback + exploration bonus}) {$R \neq 0$  {\bf and} $t>R_{length}$} {
  Compute guiding feedback $x = Env\_dist(R)$ \; 
  Store $x$ in $\mathcal{D}$ \;
  Compute $\hat{R}_{t+1}$ using SGD \cite{warnell2017deep} and $x$ \;
  }
  
  \uElseIf (\tcp*[f]{Train on past user feedback}) {$|\mathcal{D}| > 2*batchsize$}{
  $\mathcal{D}_{t+1} =$ random sample from $\mathcal{D}$ \;
  Compute $\hat{R}_{t+1}$ using SGD \cite{warnell2017deep} and $\mathcal{D}_{t+1}$ \;
  }
  
  \uElseIf (\tcp*[f]{Train on exploration bonus}) {$t>R_{length}$}{
  Compute $\hat{R}_{t+1}$ using SGD \cite{warnell2017deep} and $R^+$ \;
  }
  
  \While (\tcp*[f]{Stop autonomous exploration mode, allow direct manipulation}) {Paused}{
  	agent.get\_currentstate() \;    
  }
  
  \uIf (\tcp*[f]{Change zone state command}) {$Change\_zone$}{
  	\For{$i$ in range($n_{samples}$)}{
     Randomly sample state $s_i$ \;
     Evaluate predictiongain$(s_i) = log(\hat{p}_{t+1}(s_i)) - log(\hat{p}_t(s_i))$\;
     }
   $S_t = argmax(predictiongain(s_i))$ \;
  }
    
  \uIf (\tcp*[f]{Zone feedback computation}) {$Zone\_feedback$}{
   $S_{00} = x[0]$ and $A_{00} = Zone\_feedback$ \;
   \For{$i$ in range($R_{length}$)}{
   	\For{$j$ in range($|S|)$}{
       Take action $A_{ij}$ and observe state $S_{ij}$ \;
       Store $S_{ij}$ in $\mathcal{D}_{t+1}$ \;
    }
   }
   Compute $\hat{R}_{t+1}$ using SGD \cite{warnell2017deep} and $\mathcal{D}_{t+1}$ \;
   $\mathcal{D} \leftarrow \mathcal{D} \cup \mathcal{D}_{t+1}$ \;
  }

}
 \caption{Deep TAMER with exploration bonus and user controls for estimating $ \hat{R}() \approx R()$.}
\end{algorithm}

\newpage

\section*{Appendix E}

In a complementary work, we validated the success of our deep reinforcement learning method for human parameter exploration, by proving that guiding the \textit{Co-Explorer} better supports user creativity than simply using a standard parametric interface. We recruited 14 participants (aged between 22 and 42 years old; 3 Female and 11 Male) with varying background in sound design (such as sound engineering students, (post-)doctoral researchers in computer music, sound engineers and IT researchers/developers).

\subsection*{Procedure}

Participants were asked to explore a parameter sound space by successively using two types of interfaces: standard parametric interface, and guiding feedback to the \textit{Co-Explorer}. The succession of interfaces was set automatically by our experimental setup, which alternated each interface three times in a row (P---C---P---C---P---C). This alternation was chosen to remove bias towards user preference of an interaction type resulting from knowledge of the timbral space. Each interface was shown to participants during 5 minutes, which made the experiment last 18 minutes. The experiment was preceded by a short interface demonstration. We defined the parameter space by selecting ten parameters from a commercial VST. Participants were also asked to save an arbitrary number of varied sounds they appreciated during exploration, following their creative will.

\subsection*{Analysis}

At the end of the experiment, participants received a questionnaire in which they were asked to evaluate their experience and the interaction itself, followed by a brief and informal discussion of their experience. Participants were asked to compare the parametric interface and \textit{Co-Explorer} using scores between contrasting adjectives using a Likert scale of 1-5 (see Table \ref{tab:adj}). These adjectives were based on the user questionnaire developed by Laugwitz et al., identifying several adjectives representing criteria in classes such as perceived ergonomic quality, perceived hedonic quality (hedonic quality focuses on non-task oriented quality aspects, for example the originality of the design or the beauty of the interface) and perceived attractiveness of a product  \cite{laugwitz2008construction}. We replaced the counterpart adjective for ``understable'' by ``non-understable'', instead of ``ambiguous'' as proposed in \cite{laugwitz2008construction}. Our motivation is that we wanted to insert a negative adjective as a positive characteristic for creativity.

\begin{table}[!h]
\begin{tabular}{|l|c|l|r|}
\hline
\textbf{Classes} & \textbf{Criteria} & \multicolumn{2}{c|}{\textbf{Adjectives}}\\ \hline
Hedonic quality & Novelty & Conventional & Inventive
\\ \cline{3-4} 
 &  & Dull & Creative \\ \cline{2-4} 
 & Stimulation & Demotivating & Motivating \\ \cline{3-4} 
 &  & Boring & Exciting \\ \hline
Ergonomic quality & Perspicuity & Confusing & Clear \\ \cline{3-4} 
 & & Ambiguous & Understandable \\ \cline{2-4} 
 & Dependability & Obstructive & Supportive \\ \cline{2-4} 
 & Efficiency & Inefficient & Efficient \\ \hline
Attractiveness & & Annoying & Enjoyable \\ \hline
\end{tabular}
\caption{Contrasting adjectives used to measure and compare the experiences of guiding the \textit{Co-Explorer} and using a standard parametric interface in human parameter exploration (based on \cite{laugwitz2008construction}).}
\label{tab:adj}
\end{table}

\subsection*{Results}

We performed a 2-factor ANOVA with replication using the scores of each participant for all adjectives, for both the \textit{Co-Explorer} and parametric interface setups. We found that the \textit{Co-Explorer} had significantly better hedonic qualities than the parametric interface [$F=4.95,p<0.05$]. This may validate our hypothesis that the \textit{Co-Explorer} successfully supports user creativity in parameter exploration. Alternatively, the parametric interface proved to be more ergonomic than the \textit{Co-Explorer} [$F=22.1, p<0.001$]. This is not a surprise, as all participants reported experience in using parametric interfaces for sound design. Last and interestingly, attractiveness for both interfaces modes was not significantly different [$F=2.23,p=0.15$]. This may suggest that the \textit{Co-Explorer} did not benefit from a positive effect due to its novelty compared to the parametric interface.

\section*{Appendix F}

\begin{table}[!h]
\begin{tabular}{|l|l|}
\hline
Collaboration & \textit{(not relevant here)} \\ \hline
Enjoyment & P5 (Section \ref{subsubsec:results}, first paragraph) \\
 & P11 (Section \ref{subsubsec:coexstrat}, third paragraph) \\ \hline
Exploration & P1 (Section \ref{subsubsec:results}, second paragraph) \\
 & P12 (Section \ref{subsubsec:results}, second paragraph) \\ \hline
Expressiveness & P2 (Section \ref{subsubsec:results}, second paragraph) \\
 & P8 (Section \ref{subsubsec:results}, fourth paragraph) \\
 & P5 (Section \ref{subsubsec:coexstrat}, first paragraph) \\ \hline
Immersion & P9 (Section \ref{subsubsec:coexstrat}, first paragraph) \\
 & P5 (Section \ref{subsubsec:coexstrat}, third paragraph) \\ \hline
Results Worth Effort & P4 (Section \ref{subsubsec:results}, fifth paragraph) \\
 & P10 (Section \ref{subsubsec:results}, fifth paragraph) \\
 & P5 (Section \ref{subsubsec:coexstrat}, first paragraph) \\ \hline
\end{tabular}
\caption{Preliminary summary that maps participants' quotes in the evaluation workshop to Creativity Support Index dimensions \cite{cherry2014quantifying}.}
\end{table}

\end{document}